\documentclass[10pt, a4 paper]{article}

\usepackage{geometry}
\usepackage{graphicx}
\usepackage{titlesec}
\usepackage{caption}
\DeclareCaptionFont{mysize}{\fontsize{8}{9.6}\selectfont}
\usepackage{siunitx}
\usepackage{authblk}
\usepackage{textgreek}
\usepackage{float}
\usepackage{optidef}
\usepackage[english]{babel}
\usepackage[square,numbers,sort&compress]{natbib}
\usepackage{multicol}
\usepackage{setspace}
\usepackage[pagewise]{lineno}
\usepackage{graphics,amssymb,amsmath,epsfig,color}
\usepackage[T1]{fontenc}
\usepackage[scaled]{helvet}

\title{High Q Hybrid Mie-Plasmonic Resonances in Van der Waals Nanoantennas on Gold Substrate}

\author[ ]{Sam A. Randerson\textsuperscript{1}*}
\author[1]{Panaiot G. Zotev}
\author[1]{Xuerong Hu}
\author[1]{Alexander J. Knight}
\author[1]{Yadong Wang}
\author[1]{Sharada Nagarkar}
\author[1]{Dominic Hensman}
\author[2]{Yue Wang}
\author[ ]{Alexander I. Tartakovskii\textsuperscript{1}**}

\affil[1]{Department of Physics and Astronomy, University of Sheffield, Sheffield, S3 7RH, UK}
\affil[2]{Department of Physics, School of Physics, Engineering and Technology, University of York, York, YO10 5DD, UK}
\affil[ ]{*sranderson1@sheffield.ac.uk \quad **a.tartakovskii@sheffield.ac.uk}

\date{}

\begin{document}
\maketitle

\begin{abstract}
Dielectric nanoresonators have been shown to circumvent the heavy optical losses associated with plasmonic devices, however they suffer from less confined resonances. By constructing a hybrid system of both dielectric and metallic materials, one can retain the low losses of dielectric resonances, whilst gaining additional control over the tuning of the modes with the metal, and achieving stronger mode confinement. In particular, multi-layered van der Waals materials are emerging as promising candidates for integration with metals owing to their weak attractive forces, which enable deposition onto such substrates without the requirement of lattice matching. Here we use layered, high refractive index WS\textsubscript{2} exfoliated on gold, to fabricate and optically characterise a hybrid nanoantenna-on-gold system. We experimentally observe a hybridisation of Mie resonances, Fabry--P\'erot modes, and surface plasmon-polaritons launched from the nanoantennas into the substrate. We achieve experimental quality factors of hybridised Mie-plasmonic modes to be 19 times that of Mie resonances in nanoantennas on silica, and observe signatures of a supercavity mode with a Q factor of 263 $\pm$ 28, resulting from strong mode coupling between a higher-order anapole and Fabry--P\'erot-plasmonic mode. We further simulate WS\textsubscript{2} nanoantennas on gold with an hBN spacer, resulting in calculated electric field enhancements exceeding 2600, and a Purcell factor of 713. Our results demonstrate dramatic changes in the optical response of dielectric nanophotonic structures placed on gold, opening new possibilities for nanophotonics and sensing with simple-to-fabricate devices.
\end{abstract}

In the last decade, transition metal dichalcogenide (TMD) monolayers have attracted a large research effort owing to their direct band gap transition and useful opto-electronic properties \cite{mak2016photonics}, rendering the bulk material largely overlooked. More recently, nanoresonators utilising bulk TMDs to host Mie resonances \cite{mie1908beitrage} have gained interest, however these studies primarily focus on fabricating such structures on low refractive index dielectric substrates such as SiO\textsubscript{2} \cite{verre2019transition, green2020optical, busschaert2020transition, zotev2022transition}. A hybrid TMD-nanoantenna-on-metal system has not been thoroughly explored, with previous work being dominated by numerical simulations. One of the first to be numerically characterised was a dielectric nanowire above a metallic substrate by Oulton \textit{et al.} \cite{oulton2008hybrid}, which led to the prediction of hybrid dielectric-plasmonic modes. Since then, a variety of dielectric-metal nanodevices have been considered, from metallic nanoparticle-on-TMD monolayer systems \cite{kleemann2017strong,liu2021strong}, to hybrid nanoantennas composed of silicon and gold coaxial layers \cite{shibanuma2017efficient, sun2019enhanced}. Furthermore, dielectric nanoantennas situated several nanometers above a silver substrate have been numerically analysed by Yang \textit{et al.} \cite{yang2017low}. Their work shows hybrid dielectric-plasmonic modes \cite{yang2017low} with quality (Q) factors up to $\sim10^3$ and Purcell enhancements of $>5000$, along with strong electric field confinement in the nanoantenna-substrate gap. Shen \textit{et al.} \cite{shen2022tuning} built on this by simulating an hBN-WSe\textsubscript{2} heterostructure placed within the gap and achieving strong light-matter coupling. Experimental work has so far been limited to silicon-based nanoresonators on metallic substrates. In 2018 Xu \textit{et al.} fabricated silicon nanodisks on a gold substrate, realising an anapole mode \cite{xu2018boosting}. The anapole was used to enhance third harmonic generation by two orders of magnitude compared to the same nanodisk on an insulating substrate. Maimaiti \textit{et al.} fabricated a similar device composed of polycrystalline silicon nanodisks on a gold substrate, but with the introduction of an Al\textsubscript{2}O\textsubscript{3} spacer layer between the nanodisk and substrate \cite{maimaiti2020low}. This led to an electric field enhancement in the spacer of over 40, and a Purcell factor of 300 for a vertically oriented dipole in simulation. In experiment, they demonstrated fluorescence enhancement, and surface-enhanced Raman spectroscopy of coupled molecules with a more stable signal than solely plasmonic nanodisks. Finally, and most applicable to this study, Dmitriev \textit{et al.} fabricated silicon nanorings on gold with a layer of embedded quantum emitters between them \cite{dmitriev2023hybrid}. They observed Mie resonances from the nanoring in dark field spectroscopy, along with strong directionality of the coupled emitters normal to the substrate. \par

TMDs present an attractive alternative to silicon-based nanophotonics for generating strong mode confinement owing to their higher refractive indices \cite{munkhbat2022nanostructured}, whilst also having no absorption over large parts of the visible wavelength range \cite{munkhbat2022optical}. Additionally, one can achieve high crystalline quality of thin films (from monolayer up to $\sim500$ nm) required for fabrication of nanoantennas, through simple exfoliation from bulk TMDs \cite{novoselov2005two}. Furthermore, they can be exfoliated onto a range of other materials owing to their inherent van der Waals attractive forces \cite{mak2016photonics}. This avoids difficulties such as lattice matching requirements and growth in molecular beam epitaxy chambers, which are associated with fabricating nanoantennas from other conventional dielectrics such as GaAs and GaP. Use of TMDs thus opens new possibilities in the design and fabrication of hybrid dielectric-metallic structures of a variety of thicknesses, enabling advanced control of photonic and plasmonic resonances on the nanoscale. \par

In this study, we fabricate WS\textsubscript{2} nanoantennas on a gold substrate and characterise their optical response with experimental dark field spectroscopy, which agrees well with finite-difference time-domain (FDTD) simulations. We carefully examine the Mie and anapole resonances within such devices, and observe dramatic changes in the mode structure compared to that of nanoantennas on silica, with improved Q factors. We use both simulation, and experimental scattering-type scanning near-field optical microscopy (s-SNOM) to confirm the presence of hybrid Mie-plasmonic modes \cite{decker2017strong}, which can enhance surface plasmon-polaritons (SPPs) launched by illuminated nanoantennas. We observe a Fano lineshape for these modes, unlike the Lorentzian lineshape associated with Mie modes, therefore confirming the identification of hybrid resonances. We experimentally demonstrate that they can be easily tuned to different wavelengths by changing nanoantenna geometries. Such hybridised modes exhibit experimental Q factors up to 94, nearly a factor of 20 higher than Mie modes in nanoantennas on silica \cite{zotev2022van}, highlighting potential applications in switching and sensing \cite{lee2016active,chen2018plasmonic,miroshnichenko2010fano,limonov2017fano}. \par

We further explore strong mode coupling between a higher-order anapole mode (HOAM) and a Fabry--P\'erot-plasmonic (FPP) mode within a WS\textsubscript{2} nanoantenna on a gold substrate in both experiment and simulation. Such resonances can be tuned to realise an anti-crossing in the scattering spectra, with an experimental minimum energy splitting of 48 $\pm$ 5 meV. At the point of anti-crossing, we observe signatures of a highly confined supercavity mode \cite{rybin2017high}, with a Q factor of 263 $\pm$ 28, resulting from the destructive interference of the HOAM and FPP mode outside of the hybrid nanoantenna structure. This offers one of the first practical solutions to realising a supercavity mode in finite-sized nanophotonic devices. \par

Finally, we model novel structures that include WS\textsubscript{2} nanoantennas on top of 5 nm thick layers of hBN attached to a gold substrate. With this device, we numerically achieve strong electric field enhancement up to 2647 in the hBN spacer between the nanoantenna and gold. From this, we calculate Purcell enhancements up to 713 for a dipole polarised normal to the substrate within the hBN. This configuration offers the potential for enhancing the emission of coupled single photon emitters (SPEs), or interlayer and moiré excitons \cite{tran2019evidence,huang2022excitons} in TMD heterostructures placed within the gap. \newline
\par

\textbf{\large Results}
\par
\begin{figure}[H]
    \centering
    \includegraphics[width=\linewidth]{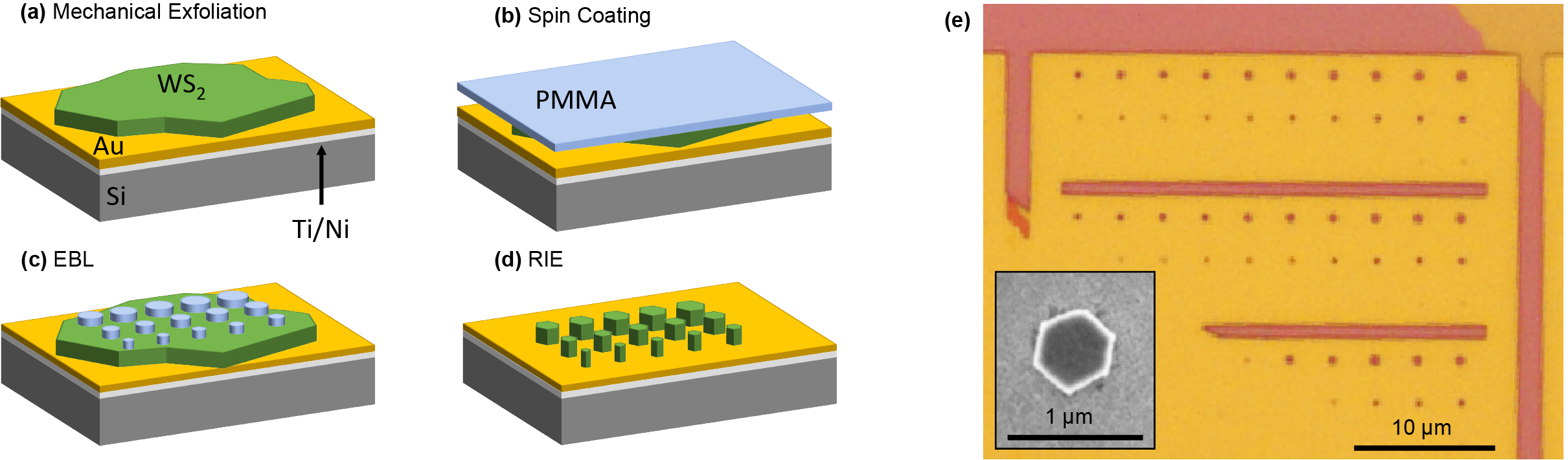}
    \caption{\textbf{WS\textsubscript{2} nanoantennas on gold fabrication and imaging.} \textbf{(a)}-\textbf{(d)} Standard nanofabrication technique of WS\textsubscript{2} nanoantennas on gold. Material thicknesses not to scale. \textbf{(e)} Optical image of nanoantennas arranged into arrays of 30 with increasing radii. Smaller nanoantennas not visible. Inset shows SEM image illustrating the hexagonal shape of the nanoantennas.}
    \label{fig.Fab}
\end{figure}

\textbf{Sample Fabrication}
We realised WS\textsubscript{2} nanoantennas on a gold substrate using well-established nanofabrication techniques (see Methods). We began by fabricating gold film substrates using two methods: electron-beam evaporation of gold onto a silicon substrate with either a titanium or nickel adhesion layer, and template-stripping \cite{ederth2000template}. We measured RMS roughness values down to 1.2 and 0.7 nm respectively, before mechanically exfoliating bulk WS\textsubscript{2} \cite{novoselov2005two} directly onto the gold. Flakes of varying thicknesses were found in each exfoliation run, measured using atomic force microscopy (AFM), before selecting those with thicknesses matching our simulations for nanoantenna fabrication. A positive resist was then spun onto our sample, before using electron beam lithography (EBL) to pattern arrays of circles of increasing radii from 100 to 400 nm. We then carried out reactive ion etching (RIE) to remove excess WS\textsubscript{2} leaving behind nanoantennas. We used an isotropic etching recipe with SF\textsubscript{6} gas (pressure and DC bias detailed in Methods) to achieve nanoantennas with hexagonal geometries. Fluorine radicals in the plasma etch the armchair axis of the WS\textsubscript{2} crystal much faster than the zig-zag axis, leaving the adjacent etched nanostructure facets at 120\textdegree{} to each other \cite{danielsen2021super,zotev2022transition,munkhbat2022nanostructured}. The true radii of the nanoantennas, as measured by scanning electron microscopy (SEM), were smaller than those in the resist pattern due to the isotropic etching process, which slightly reduces their lateral size. The gold substrate is chemically inert to SF\textsubscript{6}. This means that it acts as a natural etch stop unlike SiO\textsubscript{2}, which would be etched along with the WS\textsubscript{2}, leaving the nanoantennas on a small pedestal of substrate material. Here, the use of gold eliminates this problem. Figures \ref{fig.Fab}(a)-(d) illustrate the fabrication procedure, and Figure \ref{fig.Fab}(e) shows both optical and SEM images of the finalised nanoantennas. 
\par

\begin{figure}[H]
    \centering
    \includegraphics[width=\linewidth]{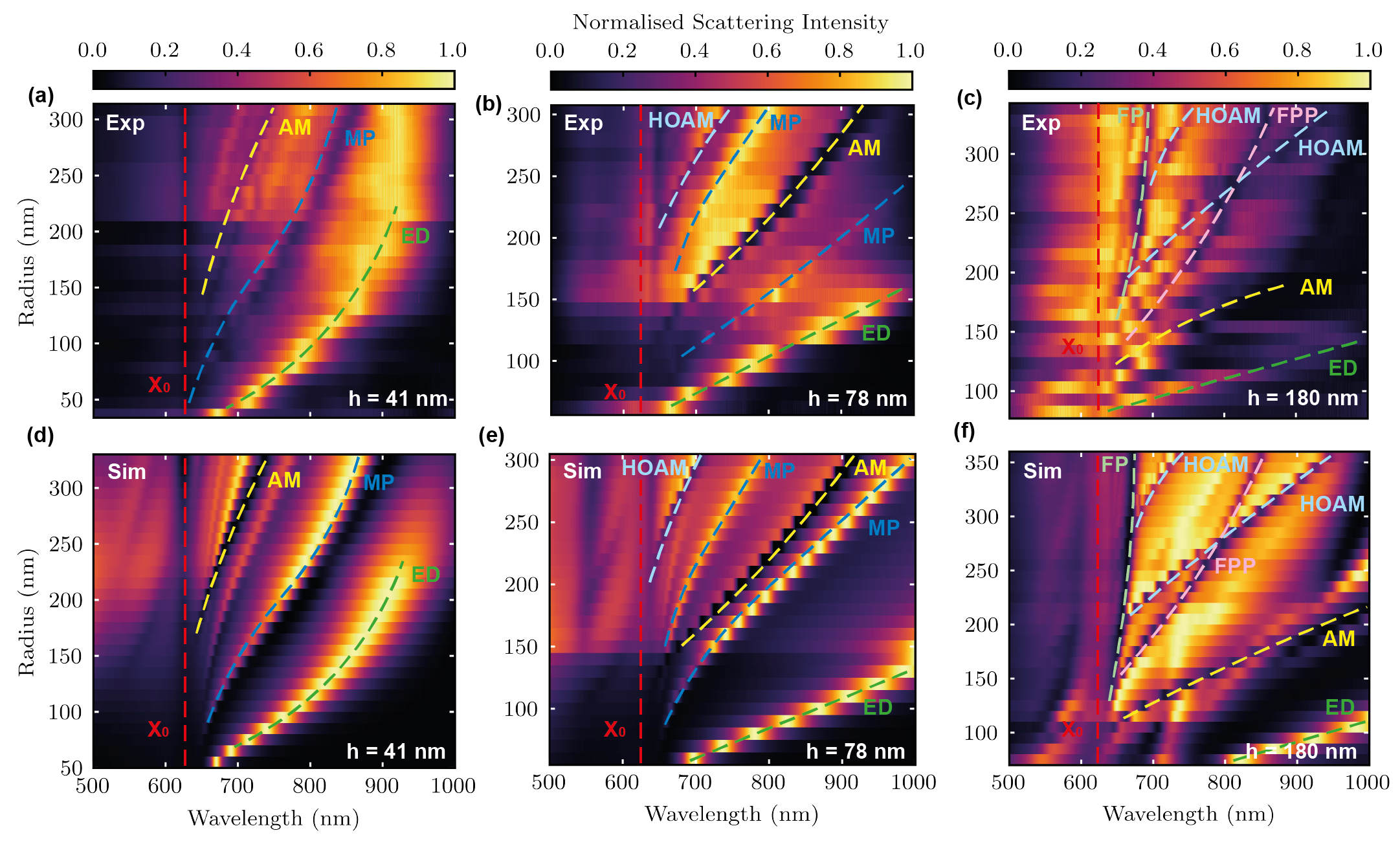}
    \caption{\textbf{Optical characterisation of WS\textsubscript{2} nanoantennas on gold.} \textbf{(a)}-\textbf{(c)} Normalised scattering cross section from experimental dark field spectroscopy of nanoantennas on gold of heights 41, 78, and 180 nm respectively. \textbf{(d)}-\textbf{(f)} Simulated normalised scattering cross section of WS\textsubscript{2} nanoantennas for the same geometries as in experiment. Left and central columns show data corresponding to single pillars (monomers), whilst right column corresponds to double pillars (dimers) with a separation of 475 nm. ED corresponds to the electric dipole mode, and AM and HOAM correspond to the anapole and higher-order anapole modes respectively. MP, FP, and FPP correspond to Mie-plasmonic, Fabry--P\'erot, and Fabry--P\'erot-plasmonic modes respectively. X\textsubscript{0} represents the WS\textsubscript{2} exciton.}
    \label{fig.Scat}
\end{figure}

\textbf{Dark Field Spectroscopy of WS\textsubscript{2} Nanoantennas on Gold}
In order to optically characterise our fabricated samples, we carried out dark field spectroscopy on individual WS\textsubscript{2} nanoantennas (monomers) on gold. We measured three different heights of nanoantennas (41, 78, 180 nm as measured by AFM) with a range of radii for each, and plot the normalised scattering intensity in Figures \ref{fig.Scat}(a)-(c). Panels (d)-(f) show the simulated scattering cross sections for the same range of heights and radii using the FDTD method, exhibiting good agreement. Note that panels (c) and (f) correspond to double nanoantennas (dimers) with a gap between them on the order of 500 nm. We do not expect a significant change in the scattering intensity between monomers and dimers, especially for such large gap sizes \cite{zotev2022transition}.

The mode structure becomes increasingly complex as we increase the nanoantenna height. In the simplest case of monomers with height 41 nm (Figures \ref{fig.Scat}(a) and (d)), we observe a variety of maxima and minima in the scattering spectra that red-shift with increasing radius. These are known as Mie resonances \cite{mie1908beitrage}, which are caused by bound charge oscillations within the WS\textsubscript{2} crystal of the nanoantennas. An example is the electric dipole mode, which leads to increased scattering in the far-field as seen by the peaks labelled ED in Figure \ref{fig.Scat}. The dark band can be identified as an anapole mode (AM), which is a destructive interference of the electric dipole mode and a magnetic toroidal mode \cite{afanasiev1995electromagnetic}, causing suppression of the scattering in the far-field \cite{miroshnichenko2015nonradiating}. The dip in scattering at a wavelength of 625 nm is due to WS\textsubscript{2} excitonic absorption, where we observe avoided crossings with the Mie modes as previously reported \cite{verre2019transition}. We also label one of the modes MP, or Mie-plasmonic, which will be discussed in greater detail in the next section. \par

We extracted the quality factor of the ED and MP mode in experiment for WS\textsubscript{2} nanoantennas on gold of height 41 nm throughout the range of radii in Figure \ref{fig.Scat}(a). A Lorentzian function was fitted to the ED mode, and a Fano curve for the MP mode to calculate each respective Q factor. We extracted maximum Q factors of 24 $\pm$ 0.6 for the ED mode of a nanoantenna with radius 48 nm, and 94 $\pm$ 5 for the MP mode of a nanoantenna with radius 125 nm. Furthermore, we compared these Q factors to those previously measured in experiment for the ED mode in WS\textsubscript{2} nanoantennas on a SiO\textsubscript{2} substrate, which reach a maximum Q factor of 5 $\pm$ 0.3 \cite{zotev2022van}. We observe a remarkable 19-fold increase in Q factor when comparing the MP mode in nanoantennas on a gold substrate, with the ED mode for nanoantennas on a SiO\textsubscript{2} substrate. \par

This behaviour is further supported by simulation. We calculated maximum Q factors from Figure \ref{fig.Scat}(d) of 61 $\pm$ 0.6 for the ED mode of a nanoantenna of radius 40 nm on gold, and 137 $\pm$ 1.6 for the MP mode of a nanoantenna of radius 100 nm on gold. By comparing these values to that of the ED mode in nanoantennas on a SiO\textsubscript{2} substrate (Q = 4 $\pm$ 0.1 \cite{zotev2022van}), we observe that the MP mode has a factor 34 higher Q factor than the ED mode for an all-dielectric system in simulation. \par

For h = 78 nm, we observe a narrowing of the electric dipole mode compared to h = 41 nm, along with a stronger red-shift with increasing radius. Additional modes such as the HOAM and two MP modes are visible in both simulation and experiment. When the height is increased to 180 nm, the mode structure becomes much more complex. Not only do we see anapole and higher-order anapole modes, but similar to previous simulations \cite{bordo2010model,landreman2016fabry,abujetas2017high,rybin2017high,bogdanov2019bound} and experiments \cite{friedler2009solid,sun2014resolving,frolov2017near,gromyko2022strong} of various dielectric nanostructures, we observe a Fabry--P\'erot (FP) mode trapped between the TMD-gold interface at the bottom of the nanoantenna, and the TMD-air interface at the top. This mode shifts very little with changing radius, as would be expected from a vertically propagating FP mode. A more rigorous definition with comparison to FP mode theory is provided in Supplementary Note 1. We also observe an anti-crossing of a HOAM and FPP mode \cite{gromyko2022strong} in both the experimental and simulated scattering spectra of the dimer nanoantennas plotted in Figure \ref{fig.Scat}(c) and (f). This occurs for nanoantennas with radii between 270 and 280 nm at a wavelength of around 800 nm, and is investigated in more detail later in this study. \par

\begin{figure}[H]
    \centering
    \includegraphics[width=\linewidth]{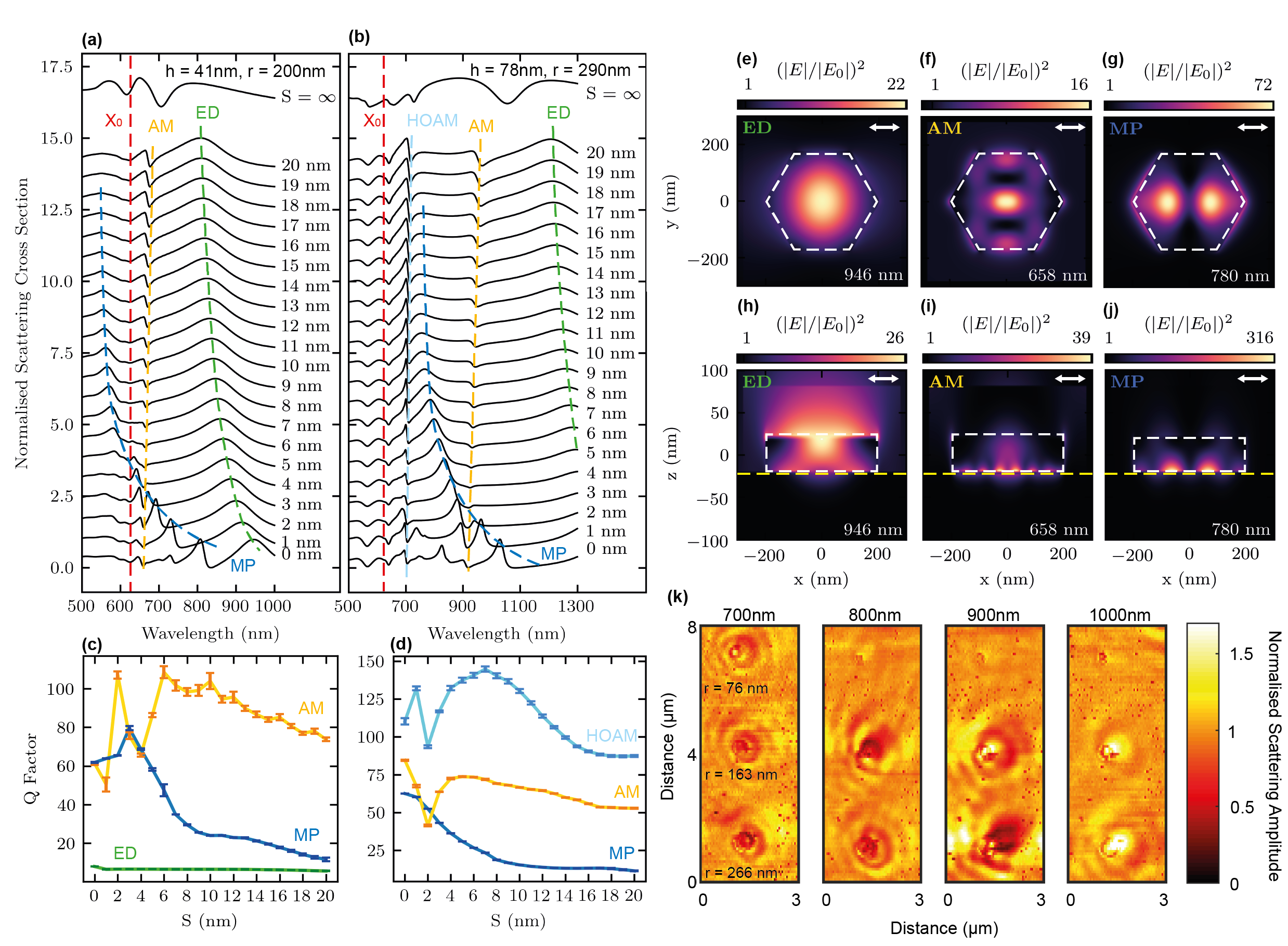}
    \caption{\textbf{Characterisation of resonant modes in WS\textsubscript{2} nanoantennas on gold.} \textbf{(a)}, \textbf{(b)} Simulated normalised scattering cross section of WS\textsubscript{2} nanoantennas with increasing distance from a gold substrate, S. Geometries are h = 41 nm, r = 200 nm, and h = 78 nm, r = 290 nm respectively. X\textsubscript{0} corresponds to the WS\textsubscript{2} exciton, ED corresponds to the electric dipole mode, MP corresponds to the Mie-plasmonic mode, AM and HOAM correspond to the anapole, and higher-order anapole modes respectively. \textbf{(c)} and \textbf{(d)} show the Q factors of each simulated resonance as a function of S. AM, MP and HOAM are fitted with a Fano curve, and ED mode fitted with a Lorentzian. Error bars correspond to error in the fitting. \textbf{(e)}-\textbf{(j)} Electric field distributions for different resonant modes within a nanoantenna of height 41 nm and radius 200 nm for S = 0 nm. Both the xy (top panel) and xz (bottom panel) views show a slice through the middle of the nanoantennas in the corresponding planes. White dashed lines represent the edges of the nanoantennas. Bottom right value corresponds to the incident wavelength. \textbf{(k)} Experimental s-SNOM scattering amplitude data for three WS\textsubscript{2} nanoantennas on a gold substrate for varying incident wavelengths as labelled above each image. All data normalised to the scattering amplitude from an area of gold free from SPP effects. Nanoantenna heights are all 41 nm with radii 76 nm (top), 163 nm (middle), and 266 nm (bottom) for comparison.}
    \label{fig.Modes}
\end{figure}

\textbf{Mie-Plasmonic Mode characterisation}
The mode structure of WS\textsubscript{2} nanoantennas in vacuum and on low-index substrates is very different compared to the case with a gold substrate (see Supplementary Note 2 for a comparison of substrates). In order to gain further insight into the origin of the modes observed in WS\textsubscript{2}-on-gold nanoantennas, we simulated a monomer of a fixed geometry positioned at a varied distance above the gold substrate. We moved the nanoantenna towards the gold in 1 nm increments, starting at 20 nm, and calculated the scattering cross section as shown in Figures \ref{fig.Modes}(a) and (b). This was done for monomer heights of 41 nm and 78 nm, and radii of 200 nm and 290 nm respectively, as studied in experiment and reported in Figure \ref{fig.Scat}. In order to characterise the Mie resonances within the nanoantennas, we performed rigorous multipole expansions of the scattered light using the open source software MENP \cite{hinamoto2021menp}. From this analysis, partial scattering cross sections attributed to the individual Mie modes can be extracted in order to assign them to their respective peaks and dips in the spectra. This was carried out for a monomer simulated in vacuum, as a homogeneous environment is necessary for the expansion. 

The green dashed line in Figures \ref{fig.Modes}(a) and (b) follows the ED mode as S is decreased from 20 to 0 nm. The peak red-shifts by 140 nm and narrows for the nanoantennas in Figure \ref{fig.Modes}(a) (h = 41 nm, r = 200 nm), therefore increasing the quality factor from 6 to 8 as shown in Figure \ref{fig.Modes}(c). We observe the electric field distribution in the xz plane of the ED mode in Figure \ref{fig.Modes}(h), where the central lobe protrudes upwards and out of the top of the nanoantenna with the introduction of the gold substrate, making the mode volume larger compared to in vacuum (see Supplementary Note 3). We therefore attribute the red-shift of the resonance to this increased mode volume. Conversely, the AM blue-shifts. Again, by considering the mode volume of the anapole resonance, as shown in Figure \ref{fig.Modes}(i), we see a confinement of the mode owing to the gold substrate. The central field maxima extends slightly less outside of the nanoantenna and into the gold than compared to the case of a nanoantenna in vacuum (see Supplementary Note 3), hence we attribute the small blue-shift to this confinement. Furthermore, a resonance peak not seen in purely dielectric systems emerges in the spectra when S is reduced to the order of 10 nm, which we name Mie-plasmonic (MP) \cite{liu2020multipole}. This peak is much sharper than that of the ED mode, and red-shifts more strongly by 159 nm from S = 4 nm to S = 0 nm for the h = 41 nm nanoantenna (Figure \ref{fig.Modes}(a)). This behaviour is also seen for the MP mode of the h = 78 nm nanoantenna. An enlarged plot of the MP mode evolution is shown in Supplementary Note 4 for clarity, where an avoided crossing with the AM and exciton is also visible. The electric field distribution of this mode in the xz plane is very strongly localised to the gold-TMD boundary shown in Figure \ref{fig.Modes}(j), which suggests that there is a contribution from plasmons, hence the naming Mie-plasmonic. This can also be observed for the ED and AM cases. However, the relative electric field enhancement at the boundary is much weaker. Interestingly, the MP mode takes a Fano lineshape in the spectra, which suggests an interference between a discrete state and a continuum of states \cite{fano2005absorption, miroshnichenko2010fano,limonov2017fano}. Both the AM and HOAM also exhibit Fano lineshapes, whereas they can be described by a Lorentzian function for the case of nanoantennas on a dielectric substrate or surrounded by vacuum \cite{svyakhovskiy2019anapole}. Supplementary Note 2 shows scattering cross sections for identical geometries of WS\textsubscript{2} nanoantennas on gold, SiO\textsubscript{2} and in vacuum for further confirmation of the different lineshapes with different substrates, suggesting that the gold substrate introduces an interference between a continuum (i.e. plasmons) and a discrete state (a Mie mode within the nanoantennas). Furthermore, the MP mode is not observed in the cases with nanoantennas on SiO\textsubscript{2} or in vacuum, and neither does it appear in previous experimental or numerical dark field studies \cite{verre2019transition, green2020optical, zotev2022transition}. This evidence further supports that there is a hybridisation of both Mie and plasmonic modes present in our TMD nanoantenna-on-gold system. \par

In addition to the previously measured Q factors from Figure \ref{fig.Scat}, we note that the AM and HOAM can reach even higher Q factors than the MP resonance when the distance between the nanoantenna and gold, S, is varied in simulation. This is shown in Figures \ref{fig.Modes}(c) and (d). For a nanoantenna of height 78 nm and radius 290 nm, the MP mode reaches a maximum Q factor of 63 $\pm${} 0.1 at S = 0 nm, whilst the AM has a maximum Q of 85 $\pm${} 0.4 likely owing to the strong lateral confinement of this type of mode. The HOAM has an even larger maximum Q factor of 145 $\pm${} 1.7 at S = 7 nm, suggesting an even stronger mode confinement. The dips in Q factor correspond to values of S where different modes cross each other in the scattering spectra, and so could be experiencing an interference effect. Whilst the Q factor of the ED mode in Figure \ref{fig.Modes}(c) only increases very slightly with the introduction of a gold substrate, we see a localisation of the mode towards the substrate in the electric field distribution in Figure \ref{fig.Modes}(h). This suggests that the ED mode also hybridises with plasmons in the gold along with the MP mode, but not as strongly. \par

To further characterise the nanoantennas and verify the mode hybridisation with plasmons, we performed s-SNOM imaging on arrays of WS\textsubscript{2} nanoantennas of height 41 nm on gold, for a range of wavelengths from 700 - 1000 nm. s-SNOM involves probing the near-field response of a sample with an illuminated AFM tip, and using interferometric techniques to resolve both the amplitude and phase of the light scattered from the tip-sample interaction region. The tip can then be scanned across the sample, as in Figure \ref{fig.Modes}(k), which shows the near-field scattering amplitude from the tip-sample interaction at each point, exhibiting the formation of ripples around the nanoantennas. Such ripples are the result of the tip illumination source interfering with SPPs on the gold, or with scattered light from features on the sample \cite{huber2008local,bozhevolnyi2001near,kaltenecker2020mono,yang2017low,yang2016near,luan2022imaging,babicheva2018near,zhang2019surface,walla2018anisotropic}. As the tip is moved across the sample, the contributions to the s-SNOM signal either interfere constructively or destructively. This produces a pattern of bright and dark fringes in the amplitude of the scattered light, which is observed as ripples. There are several methods by which this can happen, and each produce different interference patterns \cite{walla2018anisotropic,kaltenecker2020mono,huber2008local,bozhevolnyi2001near}. However, there are two prominent methods in the case of nanoantennas on gold, the first being tip-launched SPPs. As the incident light reaches the tip, it becomes strongly localised at the tip's apex. This strong near-field enhancement, and matching of the photon and plasmon wavevectors, causes tip-launched plasmons that emanate radially into the gold \cite{novotny2012principles}. Such SPPs can then reflect from nearby structures and interfere with the incident light back at the tip. The second mechanism involves SPPs launched from the nanoantennas when illuminated \cite{novotny2012principles}, which travel to the tip and interfere with the incoming light. A combination of these two effects is observed in the ripple patterns in Figure \ref{fig.Modes}(k). \par

As the incident wavelength is varied, the wavelength of the SPPs produced changes according to their dispersion relation, resulting in a change of the distance between the maxima of the ripple intensities. From Figure \ref{fig.Modes}(k), we note that the distance between the ripple maxima increases with the wavelength of the excitation laser, but is constant across nanoantenna radius. This behaviour can also be seen in the s-SNOM images of the full array of nanoantennas in Supplementary Note 5. In contrast, the amplitude of the ripples corresponds strongly to the nanoantenna size and incident wavelength. By comparing the s-SNOM data in Figure \ref{fig.Modes}(k) to the dark field spectra of the same nanoantennas in Figure \ref{fig.Scat}(a) (see Supplementary Note 5 for direct comparison), we note a correlation between the amplitude of the SPP ripples and peaks in the corresponding dark field spectra. At shorter wavelengths, such as 700 nm, we observe the highest amplitude ripples from the smallest nanoantenna (r = 76 nm), followed by the r = 163 nm nanoantenna. This agrees with our experimental dark field data, where we see a strong resonant ED mode around 700 nm for the r = 76 nm nanoantenna, and the MP mode for the r = 163 nm nanoantenna (see Supplementary Note 5). As the incident wavelength is increased to 800 nm, the r = 76 nm nanoantenna begins to resonate less in the s-SNOM images, and the r = 163 nm nanoantenna shows a stronger SPP interference pattern, which we attribute to the excitation of the MP mode. At 900 nm illumination, both the r = 163 and 266 nm nanoantennas exhibit strong SPP ripples, following the red-shift of the ED mode in the dark field data (see Supplementary Note 5). Finally, at 1000 nm illumination wavelength, the amplitude of the SPP interference pattern around all of the nanoantennas is much lower, corresponding to the dip in overall scattering in the dark field spectra (Figure \ref{fig.Scat}(a)). This observation demonstrates that van der Waals nanoantennas on gold can both scatter light to the far-field, and couple light to SPPs detectable in the near-field. Such SPPs can be further enhanced via the excitation of various Mie resonances, such as the ED and MP modes within the nanoantennas, hence providing further evidence for the coupling between Mie and plasmonic modes. \par

We do not observe SPP ripples on areas of gold far away from the nanoantennas (see Supplementary Note 6), as in this case, there are no features on the substrate to launch SPPs from. Only tip-launched SPPs are present, which travel away from the tip radially and have no edges to reflect back from. Therefore, no SPPs interfere with incident light at the tip, and so the s-SNOM image appears uniform. In Supplementary Note 6, we also show an area of the sample with pillars of resist between 25-35 nm in height on the gold, left over from the nanoantenna fabrication process. We observe negligible optical response from the resist pillars, which can be attributed to their low refractive index of 1.49 \cite{zhang2020complex}. We therefore expect tip-launched plasmons to be mostly transmitted through such structures with little reflection back to the tip. In addition, light incident on the resist pillars is not expected to lead to well confined photonic resonances unlike the WS\textsubscript{2} nanoantennas. This observation further supports our previous statement, suggesting that only the high refractive index WS\textsubscript{2} nanoantennas can launch SPPs through strongly confined, hybrid Mie-plasmonic resonances. \par

\begin{figure}[H]
    \centering
    \includegraphics[width=\linewidth]{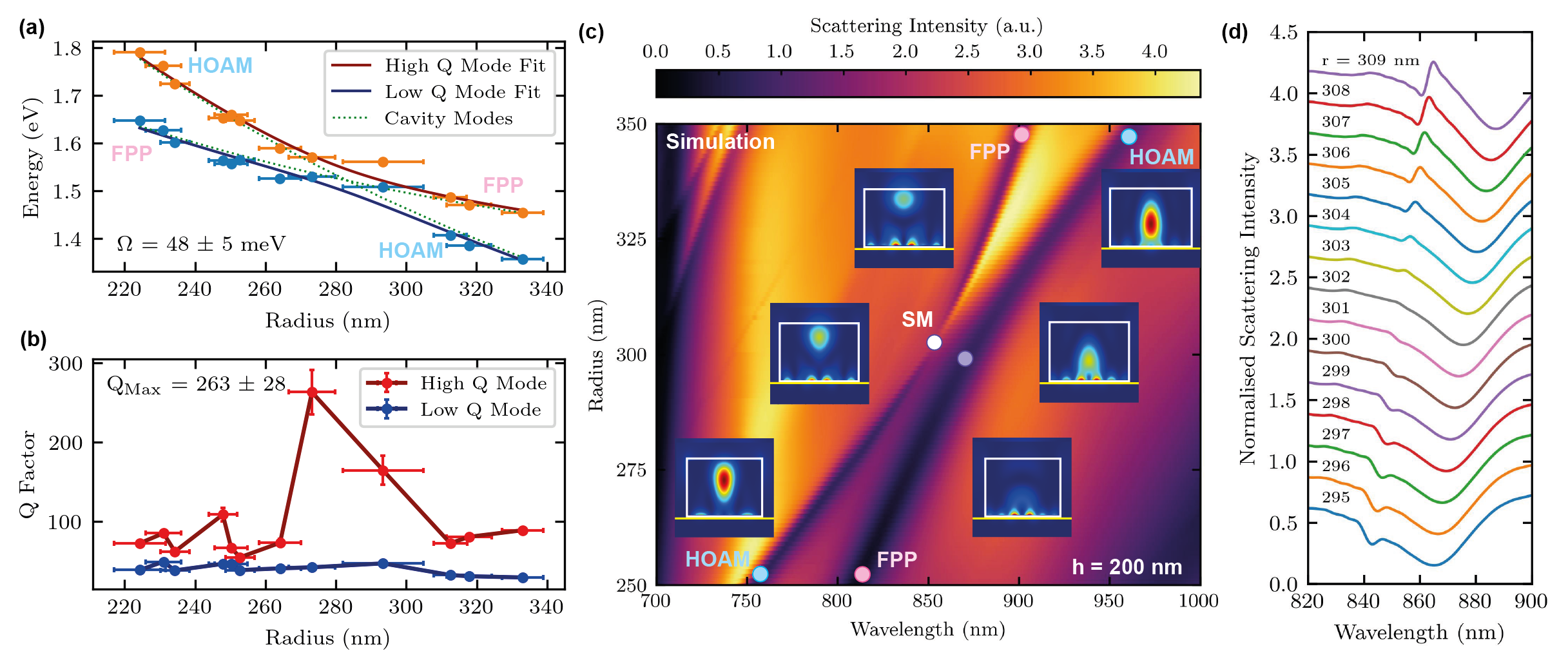}
    \caption{\textbf{Supercavity mode characterisation for a WS\textsubscript{2} dimer nanoantenna of height 180 nm and separation 475 nm, on a gold substrate.} \textbf{(a)} Experimental peak center positions of the anti-crossed HOAM and FPP mode fitted to a coupled oscillator model, yielding a minimum energy splitting of 48 $\pm$ 5 meV. Error bars represent uncertainty in the measured radii of the nanoantennas. \textbf{(b)} Quality factor of the two modes with respect to radius. Error bars represent both uncertainty in the measured radii, and error in the fits. Q factor calculated as the central wavelength of each peak divided by its respective full-width-at-half-maximum. \textbf{(c)} Simulated scattering spectra corresponding to WS\textsubscript{2} dimers of height 200 nm over a range of radii increasing in steps of 1 nm on a gold substrate. Pink and blue circles correspond to the FPP mode and HOAM respectively. White central circle denotes the position of the high Q factor supercavity mode (SM), and the purple circle corresponds to the low Q factor lossy mode. Remaining insets show electric field distributions through the center of a single nanoantenna in the xz plane at the corresponding circles in the scattering spectra. White boxes highlight the edges of the nanoantennas, and yellow lines correspond to the position of the gold substrate. \textbf{(d)} Waterfall of normalised simulated scattering spectra from \textbf{(c)} for different radii showing suppression of the high Q factor mode for the nanoantenna radius corresponding to the minimum wavelength splitting between the two modes (r = 302 nm).}
    \label{fig.Anticrossing}
\end{figure}

\textbf{Supercavity Mode in WS\textsubscript{2} Nanoantennas on Gold}
In addition to the plasmon hybridisation discussed previously, we observe signatures of a highly-confined, non-radiating supercavity mode in both experiment and simulation by tuning the radii of WS\textsubscript{2} nanoantennas on gold. This occurs as a result of the destructive interference of two different photonic modes outside of the nanoantenna, thus forming an extremely confined mode with a Q factor that, in theory, increases to infinity \cite{rybin2017high}. This is analogous to a Friedrich-Wintgen bound state in the continuum (BIC) \cite{friedrich1985interfering}, but for a finite-sized structure such as a nanoantenna. We observe signs of this in the form of an anti-crossing between the HOAM and FPP mode from Figure \ref{fig.Scat}(c). In Figure \ref{fig.Anticrossing}(a), we investigate such behaviour in greater depth, by fitting the two modes to a coupled oscillator model.

We first fitted the HOAM and FPP peaks with a double Fano formula to account for the hybridisation of the individual modes with plasmons, as detailed in Supplementary Note 7. We then extracted the peak center positions and plotted them in terms of energy against nanoantenna radius as shown in Figure \ref{fig.Anticrossing}(a). The error bars represent the uncertainty in the measured radii of each nanoantenna, with one data point at r = 293 nm showing a notably large error. We attribute this uncertainty to fabrication imperfections of this particular nanoantenna, which had a more irregular hexagonal cross-section than others when imaged with SEM, thus making the determination of its radius less reliable. The error in the fitted peak energy is negligible. The peak center positions were then fitted to a coupled oscillator model, yielding the upper and lower energy branches, shown as red and blue lines respectively. We refer to these as the high and low Q factor modes respectively. The green dotted lines represent the uncoupled HOAM and FPP mode for reference. From this fitting, we extract a minimum energy splitting of 48 $\pm$ 5 meV. This is greater than the sum of the half linewidths of the uncoupled modes (34 $\pm$ 1 meV), and hence we confirm strong mode coupling \cite{bogdanov2019bound}. \par

Furthermore, we calculate the Q factor of each peak and plot against nanoantenna radius in Figure \ref{fig.Anticrossing}(b). Whilst the low Q factor mode remains mostly constant, the high Q factor mode increases significantly at a radius of 273 nm. This radius corresponds to the closest point to the anti-crossing in Figure \ref{fig.Anticrossing}(a). We observe a maximum Q factor of 263 $\pm$ 28; an order of magnitude larger than for the ED mode reported earlier in this study. This sharp increase in Q factor along with the observation of strong mode coupling are both signatures of a supercavity mode. We expect to observe much greater Q factors by fabricating more optimised structures with changes in radii down to 1 nm, in line with our simulations detailed in Supplementary Note 7. \par

In order to better understand the mode behaviour around the anti-crossing, we performed FDTD simulations of WS\textsubscript{2} nanoantennas of height 200 nm on gold for a range of radii of 250 to 350 nm, with a much finer step in radius of 1 nm as shown in Figure \ref{fig.Anticrossing}(c). This greater height was chosen in order to red-shift the anti-crossing away from other modes in the scattering spectra to aid with fitting. The anti-crossing is clearly reproduced in the simulated spectra, and we extract an energy splitting of 37.5 $\pm$ 0.3 meV (see Supplementary Note 7), agreeing with our experimental observations. We also simulate the electric field distribution within the nanoantennas at the points marked by the coloured circles in Figure \ref{fig.Anticrossing}(c). Away from the point of anti-crossing, the FPP mode displays clear maxima and minima in the xz plane as would be expected from a Fabry--P\'erot mode confined vertically within the nanoantenna \cite{bogdanov2019bound}. However, we also see evidence of hybridisation with plasmonic modes at the gold-WS\textsubscript{2} boundary, similar to the MP modes described previously in this study. In addition, we do not observe this mode in WS\textsubscript{2} nanoantennas on a SiO\textsubscript{2} substrate in either simulation or experiment \cite{zotev2022van}. We therefore attribute this resonance to a hybrid Fabry--P\'erot-plasmonic mode as a result of reflections from, and SPPs at, the WS\textsubscript{2}-gold interface. For smaller radii the FPP mode appears mostly plasmonic, with a field maxima near the bottom of the nanoantenna. In contrast, the HOAM field distribution is strongly localised at the center of the nanoantenna, exhibiting little hybridisation with plasmons. As the radius is increased, the electric field profile of the FPP mode hybridises with that of the HOAM to form a supercavity mode labelled SM. Upon increasing the nanoantenna radius further, the HOAM returns to a similar field distribution as before the anti-crossing, with the central field maxima localised closer to the gold interface. However, the FPP mode field maxima is pushed up towards the top of the nanoantenna, whilst retaining the characteristic plasmon field distribution at the bottom. \par

We further investigate the suppression of the high Q factor mode as the nanoantenna radius is tuned. Figure \ref{fig.Anticrossing}(d) shows individual simulated scattering spectra from Figure \ref{fig.Anticrossing}(c) for a range of radii close to the anti-crossing. Whilst the low Q factor mode at higher wavelength remains mostly the same, the high Q factor mode at lower wavelength becomes almost invisible for a radius of 302 nm. This suppression of scattering corresponds to the point where the HOAM and FPP mode destructively interfere near perfectly, forming a highly confined resonance within the nanoantenna. Our simulations thus provide additional evidence to support our observation of a supercavity mode in hybrid WS\textsubscript{2}-on-gold nanoantennas in experiment. \par

\begin{figure}[H]
    \centering
    \includegraphics[width=\linewidth]{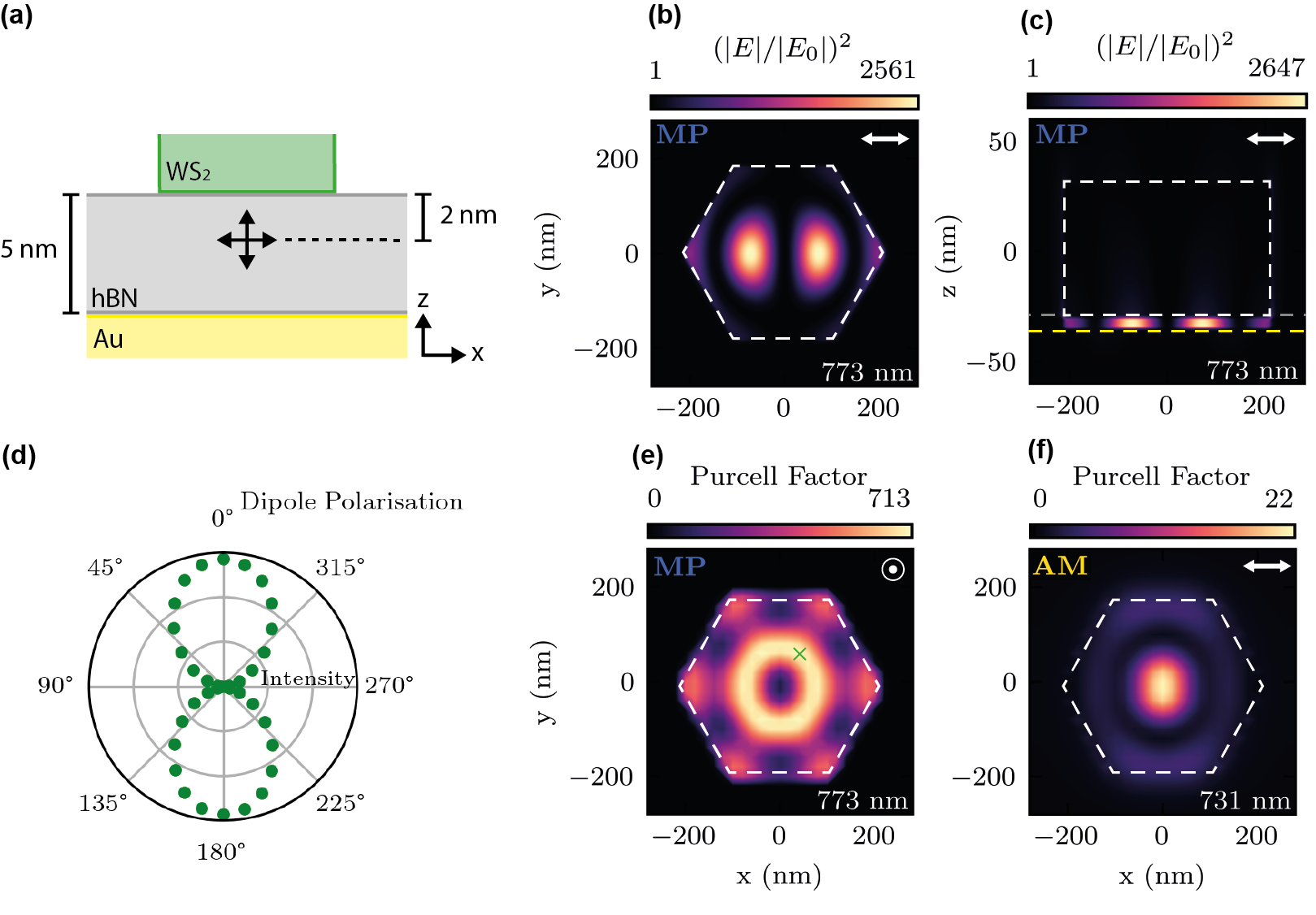}
    \caption{\textbf{Simulated electric field and Purcell factor throughout an hBN layer between a WS\textsubscript{2} nanoantenna and a gold substrate.} \textbf{(a)} Schematic of the hBN spacer showing the z position at which the dipole was placed to simulate the Purcell factor. Crossed arrows represent the two polarisations considered. Nanoantenna of height 60 nm and radius 210 nm. \textbf{(b)}, \textbf{(c)} correspond to electric field distributions within the nanoantenna-substrate gap when illuminated by a 773 nm plane wave in the xy plane, and the xz plane through the center of the nanoantenna respectively. Dashed white lines represent the edges of the nanoantenna, yellow denotes the gold-hBN boundary, and grey corresponds to the top surface of the hBN. \textbf{(d)} Polarisation dependency of the total emitted intensity of a dipole over a range of 550 - 850 nm at the position marked by the green cross in \textbf{(e)}. Dipole was rotated 360\textdegree{} in the xz plane. \textbf{(e)}, \textbf{(f)} Maps of the maximum Purcell factor for a dipole oriented perpendicular (polarised along z), and parallel (polarised along x) to the substrate respectively in the same plane as \textbf{(b)}. Dipole wavelength set to 773 nm and 731 nm respectively, corresponding to the MP and AM modes.}
    \label{fig.Efield}
\end{figure}

\textbf{Electric Field Confinement Between WS\textsubscript{2} Nanoantennas and Gold}
The strong localisation of the electric field at the TMD-metal boundary, depicted in Figure \ref{fig.Modes}(j), prompted further study into Purcell enhancement of emitters at this position. We simulated the electric field distribution within an hBN layer of 5 nm thickness between a WS\textsubscript{2} nanoantenna and the gold substrate as shown in Figure \ref{fig.Efield}(a). hBN was chosen owing to its transparency throughout the visible wavelength range \cite{nguyen2020visibility}, low refractive index of 2.2 \cite{gorbachev2011hunting}, and the presence of single photon-emitting defects, radiating at a variety of wavelengths from around 550 - 850 nm \cite{tran2016robust,jungwirth2016temperature,castelletto2020hexagonal,sajid2020single}. The results are shown in Figures \ref{fig.Efield}(b) and (c) for a nanoantenna of height 60 nm and radius 210 nm. The geometry was optimised for the maximum possible Purcell factor within the hBN layer over the wavelengths previously reported for hBN SPEs. We calculated a maximum electric field enhancement of 2647 within the hBN spacer at 773 nm wavelength (the MP mode); two orders of magnitude higher than the maximum field within the nanoantennas for the ED mode (Figure \ref{fig.Modes}(h)), and one order of magnitude higher than that of the MP mode inside a nanoantenna directly on a gold substrate (Figure \ref{fig.Modes}(j)). Based on these calculations, we spatially mapped the Purcell factor of a dipole emitter placed within the hBN (see Figure \ref{fig.Efield}(a)), mimicking an SPE. This mapping is shown in Figures \ref{fig.Efield}(e) and (f), where the dipole was oriented perpendicular (along z), and parallel (along x) to the substrate respectively.

From Figure \ref{fig.Efield}(e), we calculated a maximum Purcell factor of 713 for a dipole polarised perpendicular to the substrate at an emission wavelength of 773 nm (the MP mode), over two times greater than previously reported for silicon nanoantennas above gold for the same gap size \cite{maimaiti2020low}. Comparing to the dipole polarised parallel to the substrate in Figure \ref{fig.Efield}(f), we observed a much lower maximum Purcell factor of 22 at 731 nm (the AM). We integrated the total emitted intensity of the dipole in all directions over a wavelength range of 550 - 850 nm for varying polarisations as shown in Figure \ref{fig.Efield}(d). The dipole was located at the position of the Purcell factor maxima from Figure \ref{fig.Efield}(e), marked by a green cross, with the polarisation being rotated in the xz plane. We observed a strong sensitivity to dipole orientation within the nanoantenna-substrate gap which suggests that particular enhancement is expected in structures where the dipole is oriented vertically. An example is hetero- and homobilayer TMDs, where interlayer excitons can be observed having the electron and hole in adjacent layers \cite{louca2022nonlinear}. Therefore, we predict that by placing a TMD heterostructure within the nanoantenna-gold gap, one could achieve excitonic emission enhancement up to 32 times greater than for intralayer excitons in a similar structure, where the exciton lies in plane with the substrate. \par

\textbf{\large Conclusion}
\par
In this study, we fabricated and characterised a hybrid dielectric-metallic nanophotonic system composed of WS\textsubscript{2} nanoantennas on a gold substrate. Such TMD-based nanoantennas are easy to fabricate on metals using standard nanofabrication techniques owing to their van der Waals forces acting between the TMD thin film and the substrate, with the added benefit of gold providing a natural etch stop during RIE. We then investigated resonant Mie modes within the structures through simulation and observed excellent agreement in experimental dark field spectroscopy. We demonstrated that all the resonant modes identified can be tuned to different wavelengths simply by changing the nanoantenna radii, and that additional, higher-order modes can be introduced by increasing the nanoantenna height. Fano resonances were observed in WS\textsubscript{2} nanoantennas on a gold substrate, not present in the same structures on SiO\textsubscript{2}, which we identify as a hybridisation of Mie modes and SPPs through both simulation and experiment. These modes couple to the far-field, as measured with dark field spectroscopy, and produce SPPs detectable in the near-field with s-SNOM imaging. The SPP intensities exhibited resonant behaviour, following the red-shift of the modes upon increasing nanoantenna radii, further supporting our claims of hybridised Mie-plasmonic modes. Such hybrid Fano resonances also have high Q factors, almost 20 times higher than Mie modes in nanoantennas placed on a low-index SiO\textsubscript{2} substrate in experiment \cite{zotev2022van}, hence enabling applications in switching and sensing \cite{lee2016active,chen2018plasmonic,miroshnichenko2010fano,limonov2017fano}. \par

We further demonstrated strong mode coupling of Mie and Fabry--P\'erot-plasmonic modes within WS\textsubscript{2} nanoantennas on a gold substrate in experiment and simulation. We calculated a minimum energy splitting of 48 $\pm$ 5 meV, and with careful tuning of the nanoantenna geometry we discovered signatures of a quasi-BIC supercavity mode at the point of anti-crossing, including a significantly increased experimental Q factor of over 260, and near complete suppression of scattering in simulation. To the best of our knowledge, the use of a gold substrate reveals one of the first realistic methods of achieving a supercavity mode in van der Waals nanoantennas in experiment. \par

Finally, we observed in simulations that very strong electric field enhancement of over 2600 occurs in a nanometer scale gap between the studied WS\textsubscript{2} nanoantennas and gold substrate. For a gap filled with 5 nm of hBN, we calculated a Purcell factor of 713 for an emitter within the hBN polarised perpendicular to the substrate; 32 times higher than for parallel polarisation. This introduces opportunities for enhancing emitters placed in this nanoscale gap, such as SPEs in TMDs \cite{sortino2021bright,kumar2016resonant,srivastava2015optically,koperski2015single,kumar2015strain,luo2018deterministic,blauth2018coupling,tripathi2018spontaneous,he2015single,tonndorf2015single,palacios2017large,branny2017deterministic} and hBN \cite{tran2016robust,jungwirth2016temperature,castelletto2020hexagonal,sajid2020single}, as well as interlayer excitons in TMD bilayers \cite{louca2022nonlinear} and van der Waals heterostructures \cite{hong2014ultrafast,chen2016ultrafast,miller2017long,kunstmann2018momentum,merkl2019ultrafast,kamban2020interlayer}. \par

We have shown experimentally that merits from both plasmonic and dielectric regimes can be achieved, and that hybrid Mie-plasmonic resonances can be tuned easily for any desired application by changing nanoantenna geometries, resulting in high Q factors not previously observed in solely dielectric nanoantennas. We believe that our hybrid nanoantenna system will open up additional pathways for future nanophotonic structures and resonators, with immediate applications for single photon emitters and photoluminescence enhancement in van der Waals heterostructures. Coupling of this system to other photonic devices such as waveguides, photonic crystals, and gratings, offers near limitless combinations for using TMDs and metals together to fabricate nano-optical circuits with strong field confinement and low losses. \newline

\textbf{\large Methods}
\par
\textbf{FDTD Simulations}
In order to predict the behaviour of light within and around our nanoantennas, the software package Lumerical from Ansys was used to perform FDTD simulations.

\textbf{Scattering Simulations}
The scattering cross-sections in Figure \ref{fig.Scat} were calculated by simulating WS\textsubscript{2} nanoantennas of varying geometries on a semi-infinite gold substrate. To emulate dark field experiments as closely as possible, a total-field scattered-field (TFSF) plane wave source was used which subtracts the incident wave outside of its area of effect. This way, only the scattered light in the far-field was measured by a power monitor placed above the nanoantenna. The incident wave was set to propagate normal to the substrate and was polarised along the x axis. Anti-symmetric and symmetric boundary conditions were used along the x = 0 and y = 0 planes respectively, to reduce simulation time and memory requirements.

\textbf{Field Distributions}
To visualise the electric and magnetic field distributions within and around the nanoantennas, frequency-domain field and power monitors which perform discrete Fourier transforms (DFTs) at chosen frequencies were used. The monitors were set as 2D surfaces through the middle of the nanoantennas used in the scattering simulations along various planes, and returned the electric and magnetic field intensities normalised to the incident, vacuum wave.

\textbf{Purcell Factor Calculations}
We considered a dipole emitter placed in an hBN spacer between our WS\textsubscript{2} nanoantennas and a gold substrate to emulate an SPE. The wavelength was set to a range of 550 - 850 nm and the orientation of the dipole rotated in the xz plane to consider different polarisations. The Purcell factor was then calculated as the total integrated power of the system divided by the total integrated power of the same dipole in vacuum.

\textbf{Substrate Preparation}
The gold substrates were fabricated using either template stripping (used in the structures measured in Figure \ref{fig.Scat}(a)) or electron beam evaporation of roughly 150 nm of 99.99\% pure gold onto a silicon wafer with a 10 nm titanium (used in the structures measured in Figures \ref{fig.Scat}(b)), or nickel layer (used in the structures measured in Figure \ref{fig.Scat}(c)) to promote adhesion to the gold. These had rms roughness values down to 0.7 nm, 1.2 nm, and 2.5 nm respectively.

\textbf{TMD Exfoliation}
WS\textsubscript{2} bulk crystal from HQ-graphene was mechanically exfoliated onto the gold substrates by hand. A temperature of 105 \textdegree{}C was used to ensure good flake adhesion. Uniform thickness flakes of sizes 50 $\mu$m and upwards were recorded for patterning.

\textbf{Electron Beam Lithography}
A positive resist (ARP-9 AllResist GmbH) was first spin-coated onto the sample at 3500 rpm for 60s before heating for 2 minutes at 180\textdegree{}C. EBL was then performed using a Raith GmbH Voyager system at 50 kV accelerating voltage and 560 pA beam current. The pattern formed an array of circles of varying radii across the resist to cover several WS\textsubscript{2} flakes.

\textbf{Reactive Ion Etching}
A chemical etching recipe was used to achieve hexagonal nanoantenna geometries. Plasma etching was performed for 40 s with SF\textsubscript{6} gas at 0.13 mbar pressure with a DC bias of 50V. The armchair crystal axis of the bulk WS\textsubscript{2} was preferentially etched faster than the zigzag axis leading to 120\textdegree{} angles between them, forming hexagonal pillars \cite{danielsen2021super}. The gold substrate was etched much slower than the WS\textsubscript{2}, and so acted as a natural etch stop, leaving nanoantennas on a flat gold surface, rather than on a pedestal of substrate material. The leftover resist was then removed using warm 1165 resist remover, before bathing in acetone, followed by IPA for 5 min respectively. A final UV-ozone treatment of 20 min removed any residual organic debris.

\textbf{Dark Field Spectroscopy}
Spectroscopy involving illuminating a sample whilst rejecting the reflected light and collecting only the scattered light was achieved using a Nikon LV150N microscope with a fitted circular beam block between the illumination source (tungsten halogen lamp) and the dark field objective lens (50x with 0.8 NA). The beam block used was slightly smaller than the diameter of the beam, so that the central part was discarded and only the outer ring of light entered the objective via redirection from an annular mirror. The sample was illuminated at a high oblique angle causing light to be scattered from the sample. The vertically scattered light was then collected by the objective and passed back through the hole in the annular mirror towards a 50 $\mu$m pinhole before a fiber coupler. The pinhole ensured that only light scattered at a low angle to the normal was allowed to propagate into the 100 $\mu$m diameter core of the multi-mode fiber. Another fiber coupler then sent the beam into a free space path, where two achromatic lenses were used to minimise beam diversion along the path to the spectrometer. Finally, a single achromatic lens was used to focus the beam onto the slit of a Princeton Instruments spectrometer, where the wavelength components were separated and detected by a charge coupled device.

\textbf{s-SNOM}
Probing of the near-field scattering from our samples at the nanoscale was done using a commercial neaspec modular s-SNOM system in conjunction with a Coherent Chameleon Compact OPO-Vis pulsed laser. This technique combined a sharp AFM tip with incident radiation to strongly confine near-fields at the tip-sample interface, and measure the phase and amplitude of the scattered light. The laser was aligned onto a Platinum-Iridium (PtIr) coated cantilever tip (NanoWorld Arrow NCPt), with radius of curvature less than 25 nm, using a parabolic mirror within the s-SNOM system. The beam made a 60\textdegree{} angle with the tip, and was polarised parallel to the plane of incidence (p-polarised), to maximise the component along the tip axis. A strongly-confined near-field was generated at the tip, which then interacted with the sample as it was scanned below. Background scattering signal owing to the large spot size (few microns) compared to the tip size, was suppressed using neaspec's patented pseudo-heterodyne interferometry system. A reference beam with a phase modulation induced via an oscillating mirror was interfered with the scattered signal at the detector. This formed sidebands of frequency $n\Omega+m\Delta$, where $\Omega$ is the tapping frequency of the tip, and $\Delta$ is the modulation frequency of the reference mirror. The detector then locked in at harmonics of the tapping and sideband frequencies in order to eliminate background signal. Through using pseudo-heterodyne detection, both the amplitude and phase of the scattered light were measured simultaneously. \par
The s-SNOM measurements in this report were demodulated at either the 3\textsuperscript{rd} (Figure \ref{fig.Modes}(k) and Supplementary Note 5) or 4\textsuperscript{th} (Supplementary Note 6) harmonic of $\Omega$ and the first sideband in order to reduce the background as much as possible, whilst still keeping a good signal to noise ratio.

\textbf{\large Acknowledgements} \par
S.A.R., P.G.Z., X.H., A.J.K., S.N., D.H. and A.I.T. acknowledge support from the European Graphene Flagship Project under grant agreement number 881603 and EPSRC grants EP/S030751/1, EP/V006975/1, EP/V006975/1 and EP/V026496/1. Yadong Wang and A.I.T. acknowledge support from UKRI fellowship TWIST-NANOSPEC \linebreak EP/X02153X/1. Yue Wang acknowledges a Research Fellowship (TOAST) awarded by the Royal Academy of Engineering. S.A.R., P.G.Z., S.N. and D.H. acknowledge IT Services at The University of Sheffield for the provision of services for High Performance Computing.

\textbf{\large Author Contributions} \par
Yue Wang fabricated the gold film substrates. S.A.R. exfoliated bulk WS\textsubscript{2} onto the gold substrates and conducted AFM to ascertain flake thicknesses. Yue Wang and X.H. patterned and etched the WS\textsubscript{2} flakes into hexagonal nanoantennas using EBL and RIE. S.A.R. and Yadong Wang characterised nanoantenna radii through SEM. S.A.R. and P.G.Z. measured dark field spectra for all nanoantennas and analysed the results. S.A.R. and A.J.K. characterised the nanoantennas using s-SNOM imaging at different wavelengths. S.A.R., P.G.Z., S.N., and D.H. all contributed to FDTD simulations of the scattering spectra, electric field distributions, and Purcell factors. S.A.R., P.G.Z. and A.I.T. wrote the manuscript. A.I.T. managed the whole project.

\bibliographystyle{naturemag_modified.bst}
\bibliography{manuscript.bib}

\end{document}


\maketitle

\section*{Supplementary Note 1:}

\begin{figure}[H]
    \centering
    \includegraphics[width=\linewidth]{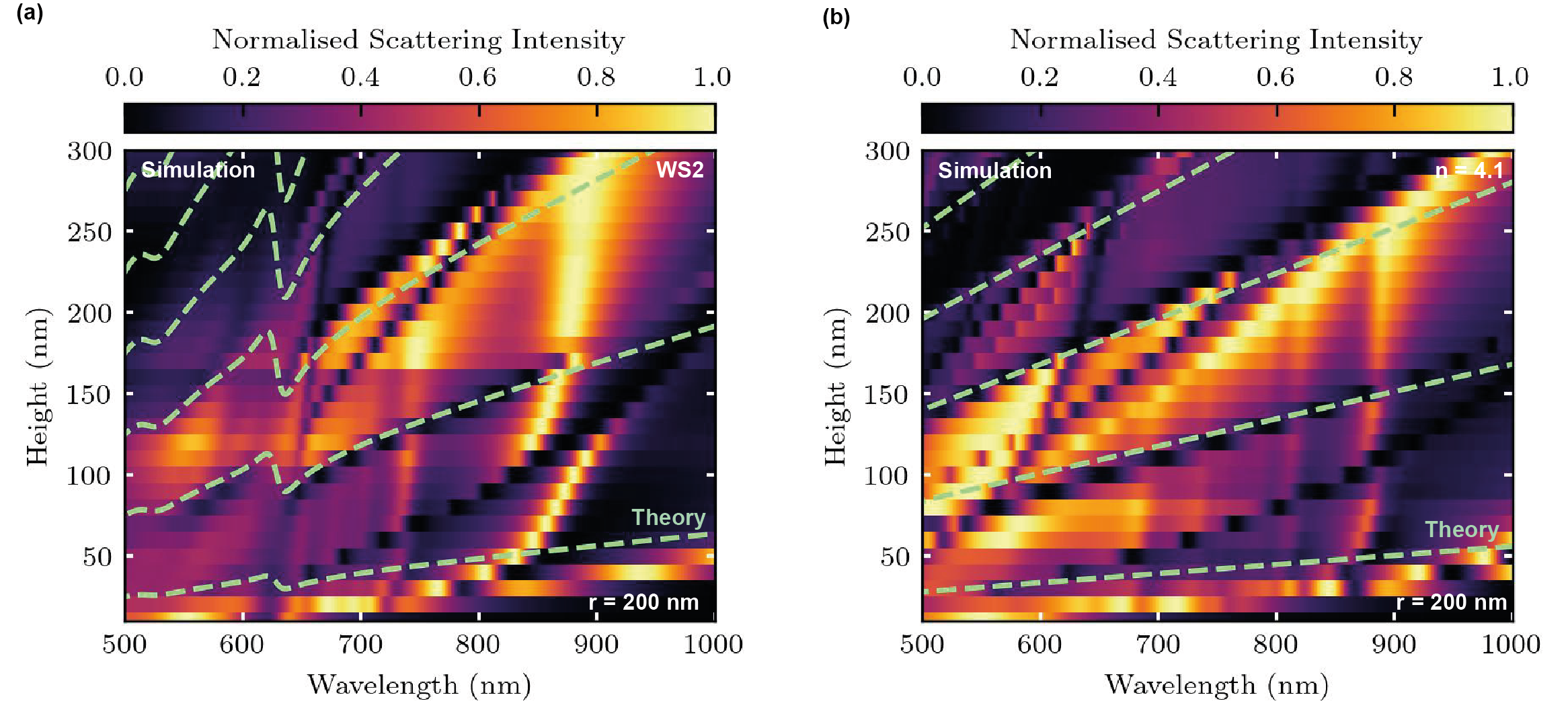}
    \caption{\textbf{Comparison of Fabry-P\'erot mode theory with FDTD data for hexagonal monomers of varying height on a gold substrate.} \textbf{(a)}, \textbf{(b)} correspond to monomers using the refractive index of WS\textsubscript{2}, and a constant refractive index of 4.1 respectively. All radii are set to 200 nm. Light green dashed lines correspond to predicted FP mode positions from theory.}
    \label{SI.FP}
\end{figure}

In order to host a Fabry-P\'erot (FP) resonance within a cavity, the round-trip phase accumulation of a wave within it must equal an integer number (m) of $2\pi$ radians. If we consider our cavity to be a dielectric nanoantenna on a gold substrate, we arrive at the following equation

\begin{equation}
\label{eqn.S1}
    2\beta H + \phi_b + \phi_t = 2m\pi,
\end{equation}

where $\beta$ is the phase constant of the wave in a nanoantenna of height $H$, and $\phi_{b,t}$ are the phase changes upon reflection from the bottom and top surfaces of the nanoantenna respectively. We assume the gold to be a perfect electrical conductor, and hence the reflection phase change at the gold-WS\textsubscript{2} boundary is $\pi$. In contrast, $\phi_t = 0$ as the wave reflects from a boundary where the outside (i.e. vacuum) is of a lower refractive index than within the nanoantenna. The phase constant can then be calculated from

\begin{equation}
\label{eqn.S2}
    \beta = \omega \sqrt{\frac{\mu\epsilon}{2} \left(\sqrt{1 + \left(\frac{\sigma}{\omega\epsilon} \right)^2} + 1\right)}
\end{equation}

where $w$ is the vacuum angular frequency, $\mu$ and $\epsilon$ are the absolute permeability and permittivity of WS\textsubscript{2} respectively, and $\sigma$ is its electrical conductivity. By then setting $H$ in Equation \ref{eqn.S1}, we can predict the wavelengths at which the total round-trip phase equals $2m\pi$, as shown in Figure \ref{SI.FP}. \par

We see good agreement of our FP mode model (dashed green lines in Figure \ref{SI.FP}) with the dark modes in our FDTD simulations for both WS\textsubscript{2}, and constant refractive index (n = 4.1) monomers on gold. For the n = 4.1 monomers, the FP modes red-shift linearly with height in the scattering spectra, and in the theory. When WS\textsubscript{2} is considered in Figure \ref{SI.FP}(a), the exciton at 625 nm causes many of the modes to become unresolvable. However, after this wavelength the mode also red-shifts linearly with increasing nanoantenna height. This is predicted by theory well, where we see an asymmetric feature around the excitonic absorption, followed by a linear increase in wavelength with nanoantenna height. Notably, the modes in the scattering spectra are blue-shifted compared to the theory. This can likely be attributed to the lateral confinement of the FP mode owing to the finite size of the nanoantennas. The FP mode model used considers an infinite plane dielectric, and so does not take into account edge effects introduced by the nanoantennas.

\clearpage

\section*{Supplementary Note 2:}

\begin{figure}[H]
    \centering
    \includegraphics[width=\linewidth]{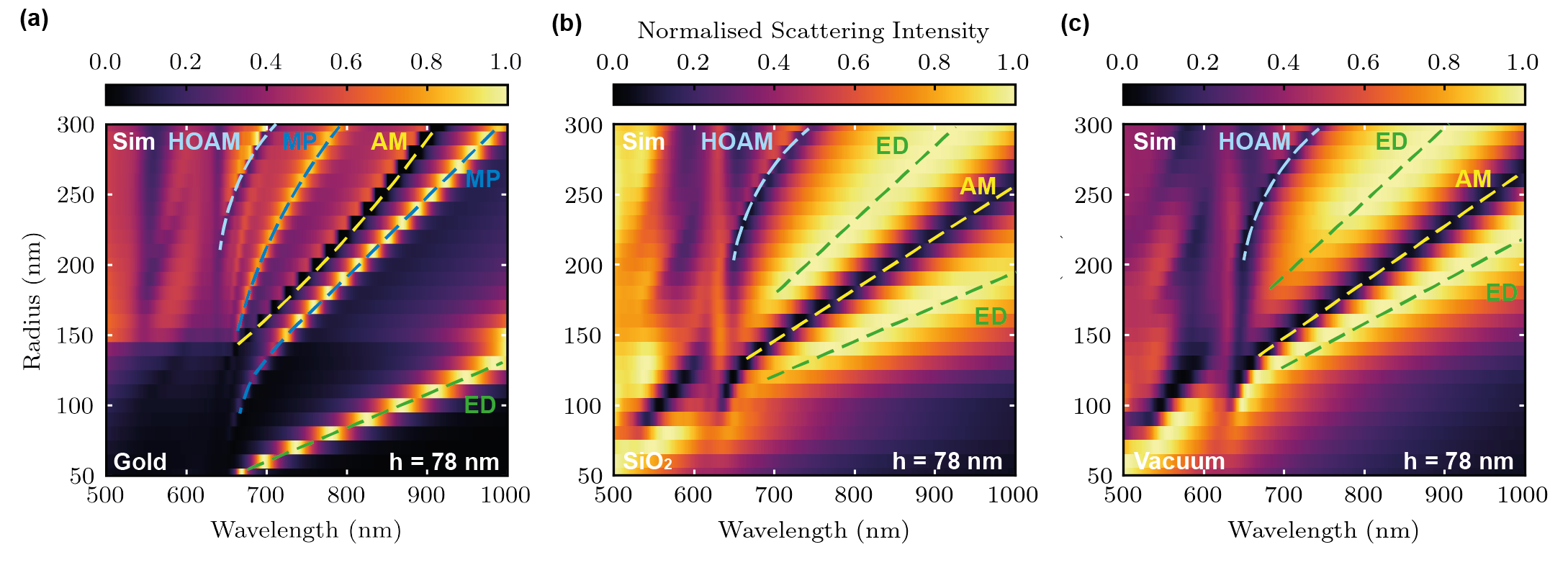}
    \caption{\textbf{Normalised scattering intensity for WS\textsubscript{2} hexagonal monomer nanoantennas of height 78 nm with different substrates in simulation.} \textbf{(a)}, \textbf{(b)}, and \textbf{(c)} correspond to gold, SiO\textsubscript{2} and vacuum substrates respectively, for the same range of radii. ED corresponds to the electric dipole mode, MP corresponds to Mie-plasmonic mode, and AM and HOAM stand for anapole and higher-order anapole mode respectively.}
    \label{SI.Scat.Comp}
\end{figure}

As shown in Figure \ref{SI.Scat.Comp}, we expect a vastly differently mode structure for WS\textsubscript{2} nanoantennas placed on a gold substrate compared to that on a SiO\textsubscript{2} substrate. The bright modes are much narrower when using gold, suggesting higher Q factor resonances. In addition, the dark mode (i.e. the anapole) exhibits a Fano lineshape when using a gold substrate, but is described by a Lorentzian in the SiO\textsubscript{2} and vacuum cases. The Fano curve suggests an interference of a discrete state and a continuum (Mie mode and plasmons), which is further reinforced by the fact that we do not see this lineshape with a dielectric substrate. Another point to note is that the AM and HOAM modes are blue-shifted when using a gold substrate compared to SiO\textsubscript{2}. This suggests a confinement of the modes, likely due to the gold substrate which reflects much of the light back into the nanoantenna unlike SiO\textsubscript{2}. In contrast, the ED mode red-shifts with the introduction of a gold substrate which can be explained by the increased mode volume seen in Figure 3(h) of the main text. WS\textsubscript{2} nanoantennas in vacuum (Figure \ref{SI.Scat.Comp}(c)) are shown as a comparison. The scattering intensity is very similar to that of the case with a SiO\textsubscript{2} substrate, owing to its low refractive index throughout the visible wavelength range \cite{rodriguez2016self}.

\clearpage

\section*{Supplementary Note 3:}

\begin{figure}[H]
    \centering
    \includegraphics[width=\linewidth]{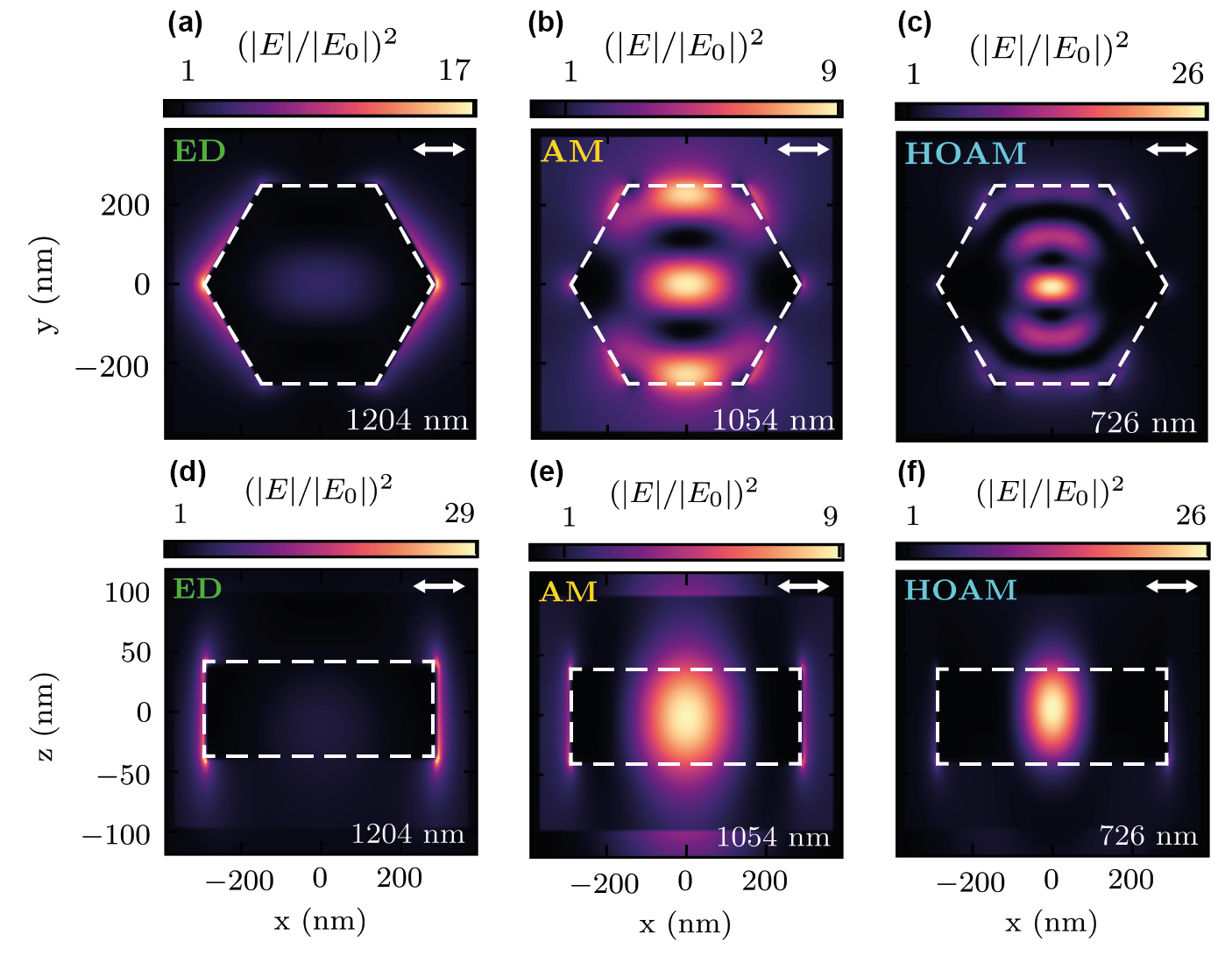}
    \caption{\textbf{Simulated electric field intensity within a WS\textsubscript{2} nanoantenna of height 78 nm and radius 290 nm in vacuum for different resonant Mie modes.} \textbf{(a)}, \textbf{(b)}, and \textbf{(c)}, correspond to the electric fields at the electric dipole mode, anapole mode, and higher-order anapole mode in a cross-section through the middle of the nanoantenna in the xy plane respectively. \textbf{(d)}, \textbf{(e)}, and \textbf{(f)} correspond to the same modes in the xz plane. Incident plane wave is polarised along the x axis. Bottom right value in each panel denotes incident wavelength.}
    \label{SI.Modes.Vacuum}
\end{figure}

The electric dipole mode for a hexagonal WS\textsubscript{2} nanoantenna suspended in vacuum is confined very strongly at, and around the vertices \cite{choi2017self} along the polarisation direction of the incoming light, as shown in Figures \ref{SI.Modes.Vacuum}(a) and (d). A central lobe is present within the nanoantenna, however the electric field intensity is much weaker than at the vertices. In Figures \ref{SI.Modes.Vacuum}(b) and (e), we note that the anapole resonance is strongly confined to within the nanoantenna boundary, with very little leakage to the environment. This is also true for the HOAM, with both resonances possessing high Q factors owing to the strong confinement. These field distributions serve as a comparison to those shown in Figures 3(e)-(j) for WS\textsubscript{2} nanoantennas on a gold substrate. Since there is no substrate in Figure \ref{SI.Modes.Vacuum}, we see that the electric field lobes are mostly symmetric about the z = 0 plane, extending equally out of the top and bottom of the nanoantenna structure. 

\clearpage

\section*{Supplementary Note 4:}

\begin{figure}[H]
    \centering
    \includegraphics[width=\linewidth]{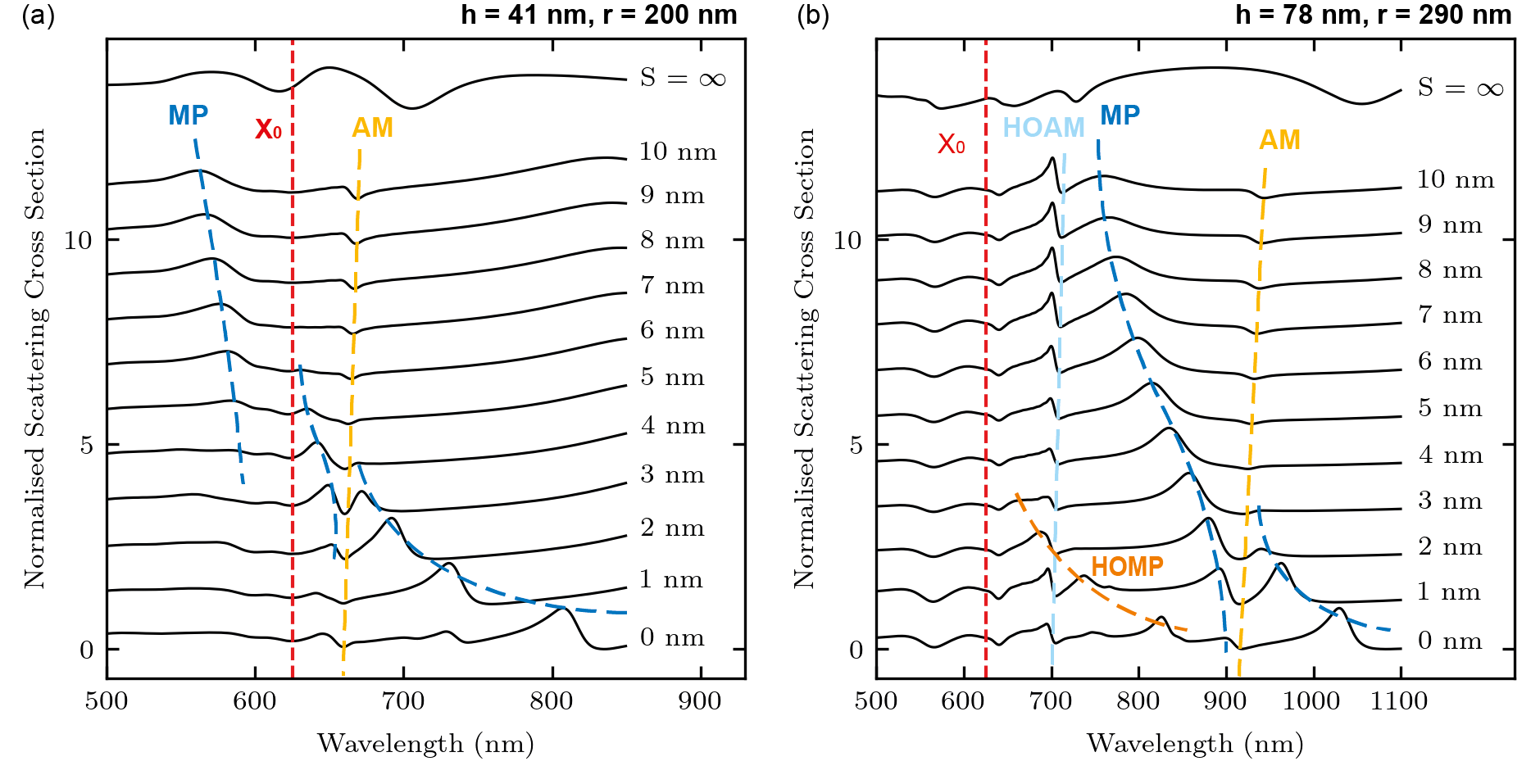}
    \caption{\textbf{Simulated scattering spectra with changing distance between a WS\textsubscript{2} nanoantenna and gold substrate, S.} \textbf{(a)}, \textbf{(b)} Enlarged scattering spectra from Figures 3(a) (h = 41 nm, r = 200 nm) and (b) (h = 78 nm, r = 290 nm) of the main text respectively for clearer visualisation of the MP mode evolution with the introduction of a gold substrate. MP, HOMP, AM and HOAM correspond to the Mie-plasmonic, higher-order Mie-plasmonic, anapole, and higher-order anapole modes respectively. X\textsubscript{0} corresponds to the WS\textsubscript{2} exciton.}
    \label{SI.Modes.Tracking}
\end{figure}

Here we show a smaller range of spectra from Figures 3(a) and (b) from the main text for S = 0 - 10 nm. A reduced wavelength range is also considered in order to illustrate the behaviour of the MP mode in more detail. We observe an anti-crossing of the MP mode with the WS\textsubscript{2} exciton in Figure \ref{SI.Modes.Tracking}(a), as seen in previous studies of purely Mie modes \cite{verre2019transition}. More interestingly, we note further avoided crossings with the anapole mode for both nanoantenna geometries in Figures \ref{SI.Modes.Tracking}(a) and (b), which is a potential indication of strong mode coupling between the two resonances. \par

In addition, when S is reduced to 1 nm, another Fano-shaped peak emerges in Figure \ref{SI.Modes.Tracking}(b) at around 740 nm wavelength. At S = 0 nm the peak center shifts to 825 nm. This strong red-shift and Fano lineshape suggests that higher-order MP modes can exist within WS\textsubscript{2} nanoantennas on a gold substrate which we term HOMP.

\clearpage

\section*{Supplementary Note 5:}

\begin{figure}[H]
    \centering
    \includegraphics[width=\linewidth]{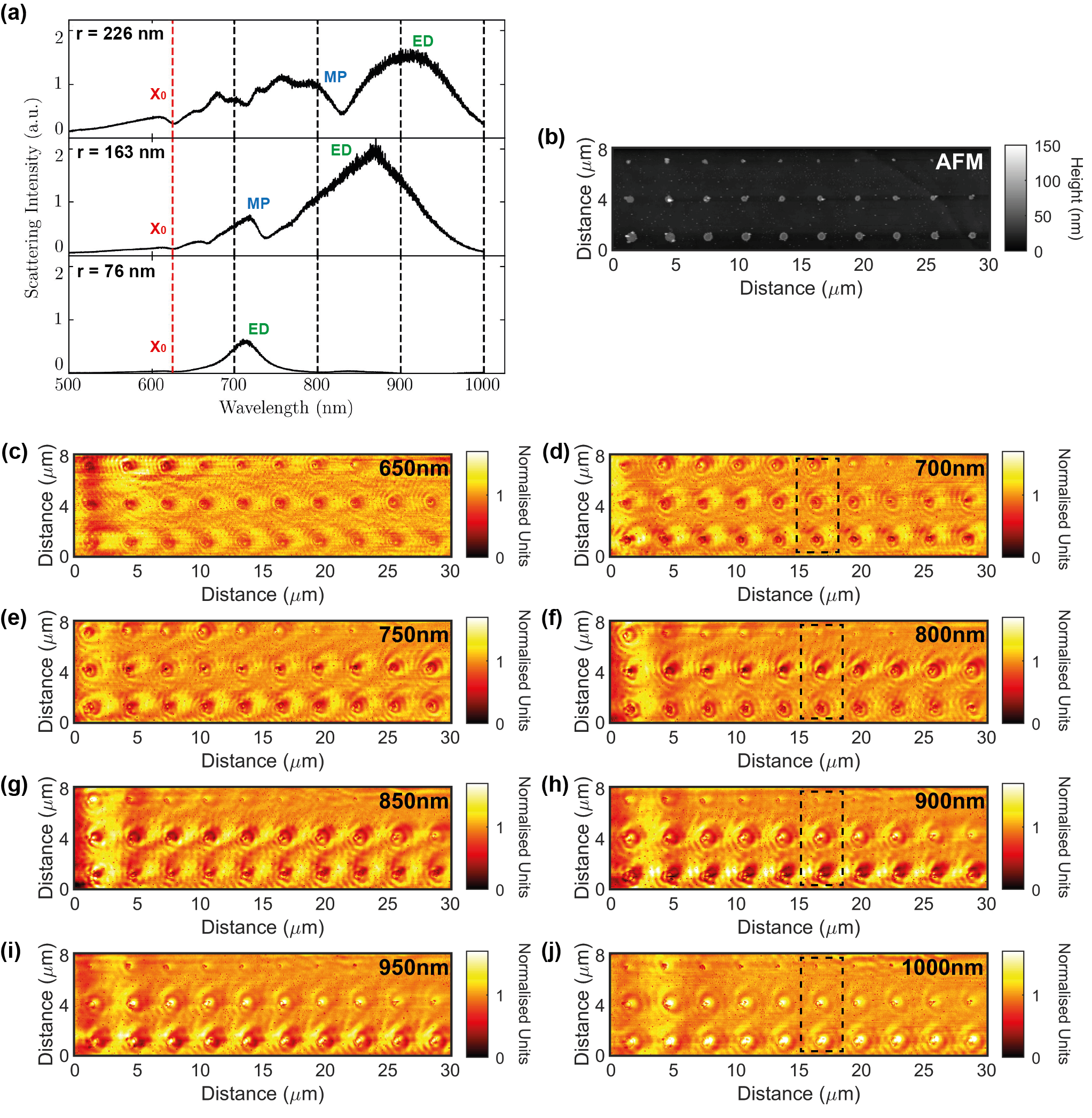}
    \caption{\textbf{Additional experimental s-SNOM data of a full array of WS\textsubscript{2} nanoantennas of height 41 nm and increasing radii on gold at different excitation wavelengths with comparison to dark field data.} \textbf{(a)} Normalised experimental dark field scattering intensity for three WS\textsubscript{2} nanoantennas on gold of height 41 nm and different radii illustrated by the black dashed boxes in Figures \ref{SI.SNOM.All.DF}(d), (f), (h), and (j). Black vertical dashed lines correspond to excitation wavelengths used in the s-SNOM measurements. Red dashed line represents the WS\textsubscript{2} exciton (X\textsubscript{0}). MP and ED correspond to the Mie-plasmonic and electric dipole modes respectively. \textbf{(b)} AFM topography image of the array of nanoantennas. \textbf{(c)}-\textbf{(j)} correspond to s-SNOM images at 650 to 1000 nm excitation in 50 nm increments. Radii increase in roughly 10 nm increments from the top right to the bottom left of the array. Black dashed boxes indicate the nanoantennas studied in Figure 3(k) of the main text.}
    \label{SI.SNOM.All.DF}
\end{figure}

We recorded s-SNOM images of WS\textsubscript{2} nanoantennas of varying radii at a range of excitation wavelengths. Most notably, we observe that the intensity of the SPP ripples follows the red-shift of Mie modes of the nanoantennas with increasing radius. This can be observed by comparing the position of the peaks in the dark field spectra of individual nanoantennas from Figure \ref{SI.SNOM.All.DF}(a), with their respective s-SNOM images (dashed boxes in Figures \ref{SI.SNOM.All.DF}(d), (f), (h), and (j)) at different wavelengths. Where we see a peak in the dark field spectra, such as for the ED and MP modes, we also observe an SPP interference pattern around the nanoantenna of higher intensity at that wavelength in the s-SNOM images. For example, at 700 nm excitation, there is a peak in the dark field spectra for the nanoantenna of radius 76 nm corresponding to the ED mode, along with a Fano-shaped MP mode for the r = 163 nm nanoantenna. By comparing this to the s-SNOM image at 700 nm in Figure \ref{SI.SNOM.All.DF}(d), we observe strong SPP ripples around the top nanoantenna in the dashed box (76 nm radius), slightly weaker SPPs around the 163 nm (middle) nanoantenna, and very little resonant behaviour for the 266 nm (bottom) radius nanoantenna, following the intensity of the modes in the scattering spectra. As the excitation wavelength is increased, the larger radii nanoantennas show higher intensity SPP patterns, whilst the 76 nm radius nanoantenna resonates with lower amplitude ripples (see Figures \ref{SI.SNOM.All.DF}(f) and (h)). Finally, at 1000 nm excitation, none of the nanoantennas show high intensity interference patterns (Figure \ref{SI.SNOM.All.DF}(j)), following the dip in the dark field scattering spectra for all three nanoantennas. This observation of the intensity of SPP interference patterns following the red-shift of Mie modes, gives further evidence to support the idea of hybridised Mie-plasmonic modes which can launch and enhance SPPs from dielectric nanoantennas on gold. \par

\clearpage

\section*{Supplementary Note 6:}

\begin{figure}[H]
    \centering
    \includegraphics[width=\linewidth]{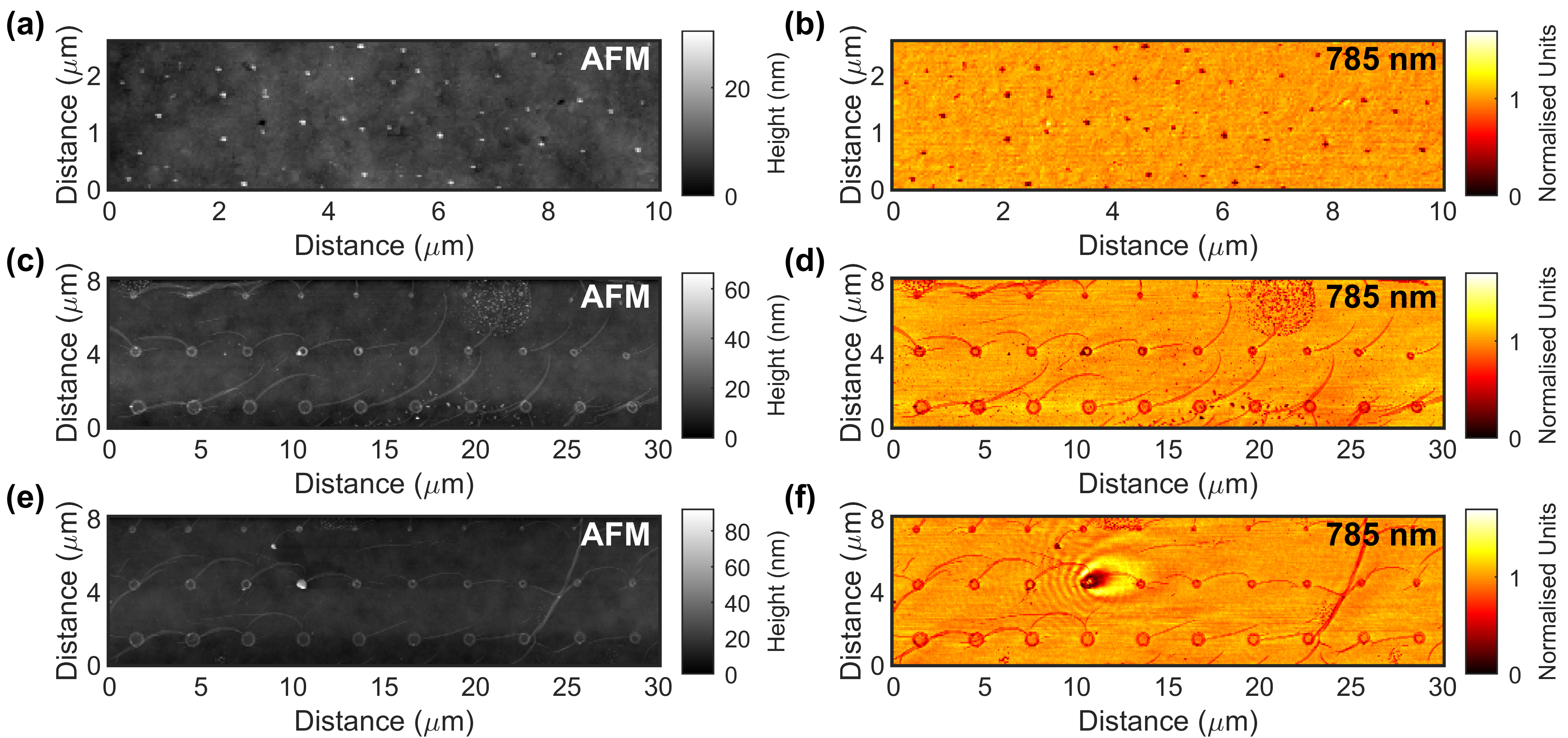}
    \caption{\textbf{AFM and s-SNOM images of gold substrate and resist pillars on a gold substrate at 785 nm excitation.} \textbf{(a)} and \textbf{(b)} correspond to AFM and s-SNOM images of the gold substrate away from any nanoantennas respectively. \textbf{(c)} and \textbf{(d)} correspond to AFM and s-SNOM images of resist pillars on gold of comparable height to the WS\textsubscript{2} nanoantennas measured in Figure 3(k) of the main text respectively. \textbf{(e)} and \textbf{(f)} correspond to AFM and s-SNOM respectively of additional resist pillars on gold with a single WS\textsubscript{2} nanoantenna remaining, surrounded by an SPP interference pattern.}
    \label{SI.SNOM.Resist}
\end{figure}

Figure \ref{SI.SNOM.Resist}(b) shows s-SNOM imaging of an area of gold substrate away from the WS\textsubscript{2} nanoantennas yielding no SPP ripple pattern as expected. Tip-launched plasmons are still present in this case, however they have no large structures nearby to reflect back from and therefore do not interference with the tip-incident light. When we introduce pillars of resist of comparable height to the WS\textsubscript{2} nanoantennas measured in Figure 3(k), we do not observe strong SPP patterns in the s-SNOM image (Figure \ref{SI.SNOM.Resist}(d)). The resist pillars do not host strong Mie resonances at visible wavelengths owing to their low refractive indices (1.49 \cite{zhang2020complex}), and so coupling to plasmons is weak. Furthermore, tip-launched plasmons are expected to transmit through such structures with very little reflection back to the tip. In Figure \ref{SI.SNOM.Resist}(f), we image an array of resist pillars with a single WS\textsubscript{2} nanoantenna in the middle. Strong SPP ripples are observed around the nanoantenna, suggesting that Mie resonances inside the structure hybridise with plasmonic resonances to enhance radially-launched SPPs. The SPP pattern is not circular in this case, owing to the incident light scattering from plasmons interacting with the tip \cite{huber2008local}. This observation further confirms the idea that only the high refractive index, WS\textsubscript{2} nanoantennas can launch SPPs through coupling with their photonic resonances.

\clearpage

\section*{Supplementary Note 7:}

\begin{figure}[H]
    \centering
    \includegraphics[width=\linewidth]{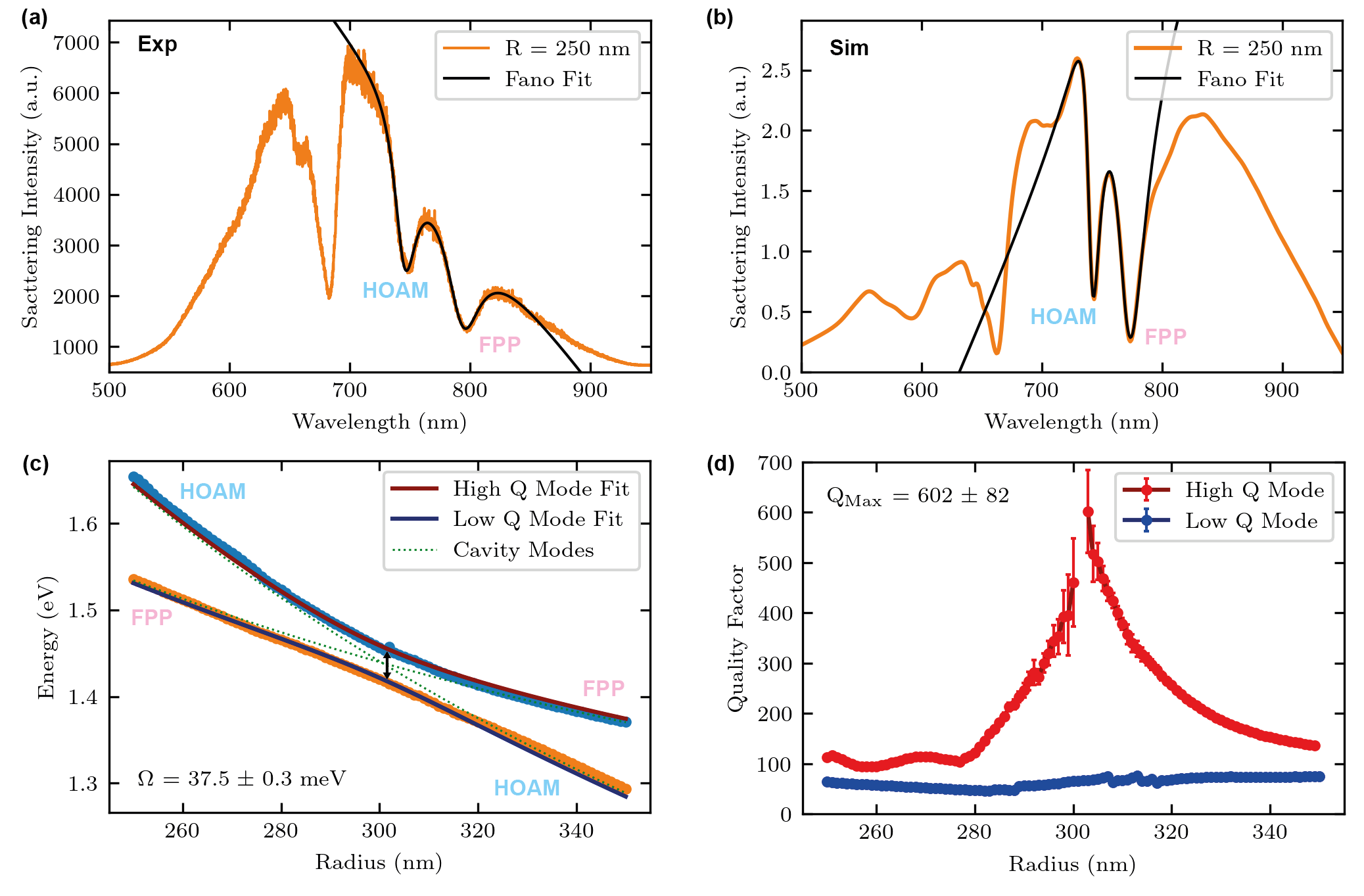}
    \caption{\textbf{Fano fits to simulated and experimental scattering spectra from WS\textsubscript{2} nanoantennas on a gold substrate.} \textbf{(a)} and \textbf{(b)} correspond to individual scattering spectra from dimer nanoantennas of height 180 nm and radius 250 nm from experiment and simulation respectively (orange lines). Black lines show double Fano curve fits to the higher-order anapole mode (HOAM) and Fabry--P\'erot-plasmonic (FPP) mode. \textbf{(c)} Simulated peak positions of the HOAM and FPP mode fitted to a coupled oscillator model, from optimised nanoantennas of 200 nm height and varying radii. $\Omega$ denotes the minimum energy splitting between the modes. \textbf{(d)} Extracted quality factor for all peaks fitted in \textbf{(c)}.}
    \label{SI.Anticrossing}
\end{figure}

In order to fit the anti-crossing observed between the higher-order anapole mode (HOAM) and Fabry--P\'erot-plasmonic (FPP) mode in Figures 2(c) and (f) of the main text, we used a coupled oscillator model. To do this, we first fitted each peak corresponding to the HOAM and FPP mode in both the simulated and experimental scattering spectra of WS\textsubscript{2} nanoantennas on gold to a double Fano curve. Examples of individual fits for single nanoantennas from both experiment and simulation are shown in Figures \ref{SI.Anticrossing}(a) and (b) respectively, corresponding to a height of 180 nm and a radius of 250 nm. This type of fit was chosen owing to the hybrid Mie-plasmonic nature of each of the modes, where we observe an interference between a resonant state and a continuum of states. To account for the two peaks, a double Fano formula was used as in Equation \ref{eqn.3},

\begin{equation}
    y = mx + y_0 - |A_l|\ \frac{(q_l + \epsilon_l)^2}{1 + \epsilon_l^2} - |A_u|\ \frac{(q_u + \epsilon_u)^2}{1 + \epsilon_u^2}
    \label{eqn.3}
\end{equation}

where

\begin{equation}
    \epsilon_{l} = \frac{x - x_{lc}}{\Gamma_{l}}, \quad \epsilon_u = \frac{x - x_{uc}}{\Gamma_u}
    \label{eqn.4}
\end{equation}

and $A_{l,u},\ q_{l,u},\ x_{l,u},\ \Gamma_{l,u}$ are the relative amplitudes, asymmetry parameters, peak center positions, and full-width-at-half-maxima of the lower and upper wavelength peaks respectively. The double Fano curves are overlaid in Figures \ref{SI.Anticrossing}(a) and (b) in black, and show good agreement with both the experimental and simulated data. We repeated this fitting process for all of the fabricated nanoantennas of height 180 nm with radii ranging from 220 to 330 nm in, on average, 10 nm increments. The peak center positions were then fitted to a coupled oscillator model as shown in Figure 4(a) of the main text. We then simulated the same sized nanoantennas but with a finer step in radius of 1 nm. We encountered difficulties in fitting owing to a broader resonance in close spectral proximity to the anti-crossing, and so optimised the nanoantenna height in order to red-shift the anti-crossing sufficiently far from the broad peak. Dimer nanoantennas of height 200 nm and radii ranging from 250 to 350 nm were then simulated. The scattering spectra were fitted using the same process as with the experimental data giving the plot in Figure \ref{SI.Anticrossing}(c), where the theoretical uncoupled HOAM and FPP mode are represented by the dotted green lines labelled cavity modes. We observe a distinct anti-crossing of the two modes, and fit them to a coupled oscillator model, yielding an upper and lower energy branch depicted by red and blue lines respectively. We refer to such lines as the high and low Q factor modes respectively, owing to their respective quality factors in the scattering spectra. From this fitting, we extract an energy splitting $\Omega$ of 37.5 $\pm$ 0.3 meV, which is greater than the sum of the half linewidths of the HOAM and FPP mode away from the closest point to the anti-crossing, hence confirming strong mode coupling. \par

Furthermore, upon plotting the Q factors of each of the branches against nanoantenna radius as in Figure \ref{SI.Anticrossing}(d), we see that the low Q factor mode remains approximately constant, whereas the high Q factor mode peaks significantly for a radius of 302 nm. We calculate a maximum Q factor of 602 $\pm$ 82 for a nanoantenna radius of 303 nm, with the high error owing to the suppression of the peak in the scattering spectra. This can be seen in Figure 4(d) of the main text, where the high Q peak becomes difficult to fit in the range 301 - 303 nm as a result of the destructive interference between the two photonic modes. This behaviour is signature of a Friedrich-Wintgen bound state in the continuum (BIC) \cite{friedrich1985interfering}, and leads to a highly confined mode with a theoretically infinite Q factor for a nanoantenna with radius corresponding to the lowest energy splitting between the modes. This is supported by our results in Figure \ref{SI.Anticrossing}(d), which shows a Q factor increasing to infinity for a radius of 302 nm, which is also the closest point to the anti-crossing in Figure \ref{SI.Anticrossing}(c). Since an infinite Q factor is, by nature, impossible to measure; we attribute this mode to a quasi-BIC from our simulated data. In the context of finite-sized nanoantennas, we term this a supercavity mode \cite{rybin2017high}.

\clearpage

\bibliographystyle{naturemag_modified.bst}
\bibliography{supplement.bib}